\documentclass[12pt]{article}

\def\+{{+\!\!\!+}} 
\def\pp{\mbox{\tiny${}_{\stackrel\+ =}$}}

\def\d{\partial}

\def\P{\Phi}

\def\r{\rho} 
\def\l{\lambda} 
 
\def\L{\Lambda}

\def\pmb#1{\setbox0=\hbox{#1}%
\kern.0em\copy0\kern-\wd0 
\kern-.04em\copy0\kern-\wd0 
\kern.08em\copy0\kern-\wd0 
\kern-.04em\raise.0433em\box0 }         
 
\def\half{\frac{1}{2}}

\newcommand{\nc}{\newcommand} 
\nc{\beq}{\begin{equation}} 
\nc{\eeq}[1]{\label{#1}\end{equation}} 
\nc{\ber}{\begin{eqnarray}} 
\nc{\eer}[1]{\label{#1}\end{eqnarray}} 
\nc{\pek}[1]{\cite{#1}} 
\nc{\enr}[1]{(\ref{#1})} 
\nc{\kal}[1]{{\cal{#1}}} 
\nc{\dott}{\;\cdot\;} 

\def\0 {\nonumber}

\begin{document} 
\setcounter{page}{0}
\newcommand{\inv}[1]{{#1}^{-1}} 

\makeatletter
\@addtoreset{equation}{section}
\makeatother
\renewcommand{\theequation}{\thesection.\arabic{equation}}


\newcommand{\be}{\begin{equation}} 
\newcommand{\ee}{\end{equation}} 
\newcommand{\bea}{\begin{eqnarray}} 
\newcommand{\eea}{\end{eqnarray}} 
\newcommand{\re}[1]{(\ref{#1})} 
\newcommand{\qv}{\quad ,} 
\newcommand{\qp}{\quad .} 
\begin{titlepage} 
\begin{center} 
                                
                                \hfill   hep-th/0405085\\ 
                                \hfill   UUITP-09/04\\
                                \hfill   HIP-2004-17/TH\\
                                \hfill   CPHT-RR 018.0504\\
                                \hfill   LPTHE-04-08\\
\vskip .3in \noindent 
 
 
{\large \bf{Generalized complex manifolds and supersymmetry}} 
\vskip .2in 
 
{\bf Ulf~Lindstr{\"o}m$^{ab}$,~Ruben~Minasian$^c$,~Alessandro~Tomasiello$^c$~and~Maxim~Zabzine$^{de}$}\\ 
 
\vskip .05in 
 
 

$^a${\em\small Department of Theoretical Physics\\ 
Uppsala University, Box 803, SE-751 08 Uppsala, Sweden} \\
\vskip .02in
$^b${\em\small HIP-Helsinki Institute of Physics\\
P.O. Box 64 FIN-00014 University of Helsinki, Suomi-Finland}\\
\vskip .02in 
$^c${\em\small Centre de Physique Th{\'e}orique, Ecole Polytechnique\\
91128 Palaiseau Cedex, France} \\
\vskip .02in 
$^d${\em\small LPTHE, Universit{\'e} Pierre et Marie Curie, Paris VI\\
 4 Place Jussieu, 75252 Paris Cedex 05, France}\\
\vskip .02in 
$^e${\em\small Institut Mittag-Leffler,
Aurav{\"a}gen 17, 
S-182 62 Djursholm, Sweden}

 
\vskip .5in 
 
\vskip .1in 
\end{center} 
 
\begin{center} {\bf ABSTRACT }  
\end{center} 
\begin{quotation}\noindent  
We find a worldsheet realization of generalized complex geometry,
a notion introduced recently by Hitchin which
interpolates between complex and symplectic manifolds. The
two--dimensional model we construct is a supersymmetric relative of the
Poisson sigma model used in context of deformation quantization. 
\end{quotation} 
\vfill 
\eject 
  

\end{titlepage}

\section{Introduction}


The recently developed notion of generalized complex geometry naturally
extends and unifies complex and symplectic geometries, in general
interpolating between the two \cite{Hitchin, Gualtieri, DH}\footnote{In fact before
 the Hitchin's work \cite{Hitchin} the algebraic aspects
  of a generalized complex (K\"ahler) geometry
 has been discussed in the physics literature \cite{Kapustin:2000aa}.}. 
There have been many hints that this geometry should be relevant to string
theory.  
In this paper, we realize this expectation from a world--sheet
perspective. 

The reasons to believe that generalized complex geometry should fit
naturally in string theory basically all stem from
the fact that the formalism puts the tangent $T$ and the cotangent
bundle $T^*$ on the same footing, considering pairs $(v,\xi)$ in
$T\oplus T^*$. The basic objects of the formalism,
generalized (almost) complex structures, are endomorphisms of this
bundle, and admit an action 
not only under diffeomorphisms but also under a two--form. As we will
see, this action is essentially a change in the string theory $B$--field.
A related remark is that the structure group of this bundle is SO$(d,d)$,
which indicates a relation to  the string theory T--duality group. This is
 strengthened by the interpolation between complex and symplectic
geometry, which are mirrors in string theory.

The formalism has in fact already found an application recently, from a
perspective different from the one in this paper. 
A mirror symmetry transformation was proposed in \cite{fmt} for manifolds of
SU(3) structure, generalizing the case of Calabi--Yau manifolds with
NS flux, considered in \cite{vafa,glmw}.
As it turns out, mirror symmetry can be expressed naturally
in terms of the $T \oplus T^*$ formalism.


In \cite{fmt}  mirror symmetry is expressed as an exchange of
two {\it pure spinors}. These objects appear in many disparate
contexts, depending on which Clifford algebra one is considering. 
In case of Clifford$(9,1)$ they play a role in Berkovits
superstring \cite{berkovits}, and in the case of Clifford$(d)$ 
they can be used to define twistor spaces for manifolds of dimension $d$ \cite{LM}.
In the present paper the relevant spinors are those of the
Clifford algebra naturally built on $T \oplus T^*$, which is
Clifford$(d,d)$.  The same way as usual Clifford$(d)$
spinors can be realized in terms of $(0,p)$ forms on an almost complex
manifold, Clifford$(d,d)$ spinors can be realized as formal sums of
forms of mixed degrees. Pure spinors are then those which have a
stabilizer of maximal dimension, which can be translated into an
algebraic condition that we will review later.  On a SU$(d)$ structure
manifold we can give two prototypical
examples, which are also those exchanged by mirror symmetry:
the $(d/2,0)$ form $\Omega$ and an exponential of the two-form, $e^{i J}$. 

These pure spinors play a role in the $T\oplus T^*$ formalism: there
is a correspondence between generalized complex
structures and pure spinor lines. These two complementary pictures of the
formalism can be seen as the two complementary pictures of string
theory -- from the world--sheet and from the supergravity point of
view. Pure spinors naturally emerge in the low--energy context, in
which, in particular, the above mentioned mirror symmetry proposal was
formulated. In this paper, we
are going to see how generalized complex structures emerge from the
world--sheet point of view. 

Another aspect which is taken into account naturally by the formalism
is the following. Mirror symmetry was defined in \cite{fmt} in the
class of manifolds of SU(3) structures. This was however a
simplification. Mirror symmetry as defined
in \cite{fmt} is inspired by T--duality along three directions. In
certain cases, essentially when the $B$--field has more than one leg
along the T--dualized directions,\footnote {For
simplicity, this  case was not considered in \cite{fmt}.} the result of T--duality only makes
sense as a ``non--geometric'' background. Usually, one thinks geometric quantities are sections of
bundles associated with the frame bundle: they transform from chart to
chart under diffeomorphisms. In string theory, the symmetry group is
larger than diffeomorphisms. One can indeed use the SO$(d,d)$
invariance mentioned above. Then, there may exist more
general SO$(d,d)$-valued transition functions, apart from the 
the usual Diff-valued ones. This will for example mix metric and
$B$--field, making them not well--defined separately. In this
situation one speaks of a non--geometrical
background. This possibility has been emphasized in many papers;
Scherk--Schwarz compactifications
are for example of this type, and also the 
 ones in \cite{hmw, Flournoy:2004vn}.\footnote{Our interest in these matters owes much to a
  conversation with S.~Hellerman, who also made the above remark about
  non--geometrical mirror symmetry.}
 Let us
also emphasize that we are not assuming the existence of global isometries,
and not doing T--duality. SO$(d,d)$ only appears as a structure group.

 Hopefully, the structure described above will allow  the formulation
of mirror symmetry using pure spinors to be extended to 
``non--geometrical'' situations.


 The present paper realizes
SO$(d,d)$ covariance and describes generalized complex geometry. The
idea is simple and is introduced in a paper by
one of the present authors \cite{Lindstrom:2004eh}. 
The usual sigma model only contains fields in
$T$, the images under the differential of the map $X$ from the
world--sheet to the target space. It does not contain objects in $T \oplus T^*$. 
A related 
 model is the Poisson sigma model which does contain fields both in $T$ and $T^*$,
and was used in the context of deformation quantization \cite{cf}. 
We mimic the structure of the Poisson sigma model for the usual one. 
We  double the number of degrees of freedom introducing 
new fields $\eta$ valued in the tangent $T^*$, and write
an action for these $2d$ fields classically equivalent to the usual sigma model.

A difference between the two actions (the
first--order one and the usual second--order sigma--model) is that while a
second-order action is fully
determined by the metric and a closed 3-form $H$, a first order action
needs a section $E$ of the $O(d,d)$ bundle $T\oplus T^*$. Any two such
first-order actions are equivalent, i.e., lead to the same set of
equations of motion, as long as they are transformed into each other by an
action of a closed 2-form $b$.
This puts the $b$-transform on equal footing with
diffeomorphisms. This already captures some features of
the formalism of generalized complex structures. 


More differences between the first and second order forms show up when we try to 
supersymmetrize the action. We
are used to the idea that requiring the action to be
supersymmetric constrains the target space geometry. In the second--order action 
this does not involve a generalized
complex structure, only complex structures. In
this paper we 
analyze the conditions under which the first order action has additional supersymmetries. 
 This was done for $N=(2,2)$ supersymmetry in \cite{ Lindstrom:2004eh} going partly
 on-shell and is done here completely off-shell for $N=(2,0)$.
Given
what we already mentioned, it is fair to expect the appearance of
generalized complex structures. 


In this paper we study models with $N=(2,0)$
supersymmetry (in the absence of boundaries), and find a 
generalized complex geometry. We consider three different cases, and the 
realization of  the generalized complex geometry depends on the details.
In particular, for a special case, at algebraic level 
we recover the $N=(2,2)$ geometry discovered in \cite{Gates:nk}\footnote{
 See also \cite{Lyakhovich:2002kc}, for recent developments.}.
The first order action serves as a basis for T-duality. Since
T-duality mixes the right 
and left sectors \cite{hassan}, this form of the action probes all models related by
such transformations.

There are many directions in which the present work might be
extended. An obvious one is the inclusion of boundaries. In particular 
this may clarify the cases discussed in \cite{Lindstrom:2002jb, Lindstrom:2002vp} for which a geometrical interpretation is lacking. 
Further, in
topological models, the relevance of generalized complex structures
has been demonstrated in \cite{kapustin} (see also \cite{grange}).
It would be interesting to twist the physical model discussed in this paper to
reproduce those results in a more general setting.

The structure of the paper is as follows. In sec. 2 we introduce
the first-order action and some notation. A brief review of generalized
complex geometry is given in sec. 3. We phrase the integrability conditions in 
local coordinates. Sec. 4 contains a
discussion of the topological model. 
It represents the most general geometric situation. As the
$T \oplus T^*$ formalism allows for twisting with 3-form $H$ it is natural to
examine  the twisted construction in our context as well. This is done in
sec. 5 where we discuss the WZ-term. The $(2,0)$ sigma model is
presented in section 6 and the geometry of the target space is discussed.
Finally, we gather the most
technical part of our computations, namely the closure of the
supersymmetry algebra, in an appendix.


\section{First order actions}

 In this Section we describe the class of two dimensional models which are relevant 
 for our discussion.

 We start by introducing the standard bosonic sigma model. This model has a single bosonic real field, $X$.
 $X$ is a map from a two-dimensional world-sheet $\Sigma$ (without a boundary) to
  a manifold ${\cal M}$ equipped with a metric $g_{\mu\nu}$ and a closed three form $H_{\mu\nu\rho}$.
 The action of the model is
\beq
 S = \frac{1}{2}\int \,\, g_{\mu\nu}(X) dX^\mu \wedge *dX^\nu + B_{\mu\nu}(X) dX^\mu \wedge dX^\nu 
\eeq{actiondefbos}
 where $H=dB$ on some patch. Although $B$ is used to write the action (\ref{actiondefbos}) down
 the theory depends only on the three form $H$. 

 We introduce a globally defined two-form $b_{\mu\nu}$ on ${\cal M}$. Then we can define
 the following tensors
\beq 
 E_{\mu\nu} = g_{\mu\nu} + b_{\mu\nu},\,\,\,\,\,\,E^{\mu\lambda} E_{\lambda\nu}=\delta^\mu_\nu,
\,\,\,\,\,\,G^{\mu\nu} =\frac{1}{2}(E^{\mu\nu} + E^{\nu\mu}),\,\,\,\,\,\,
 \theta^{\mu\nu}=\frac{1}{2}(E^{\mu\nu} - E^{\nu\mu})
\eeq{definotenrp}
 If $g$ is a Riemannian metric then $G$ is also a Riemannian metric. We can introduce a new field 
$\eta$ which is a differential form on $\Sigma$ taking values in the pull-back by $X$ of the cotangent bundle of 
 ${\cal M}$, i.e. a section of $X^*(T^*{\cal M})\otimes T^*\Sigma$. 
There exists a first order action \cite{Baulieu:2001fi, Lindstrom:2004eh} 
\beq
 S= \int \,\, \eta_\mu \wedge dX^\mu + \frac{1}{2} \theta^{\mu\nu} \eta_\mu \wedge \eta_\nu +
 \frac{1}{2} G^{\mu\nu} \eta_\mu \wedge *\eta_\nu + \frac{1}{2} (B -b)_{\mu\nu} dX^\mu \wedge dX^\nu ,
\eeq{firstorderact}
 which is equivalent to (\ref{actiondefbos}) upon the integration of $\eta$. Following the terminology 
 proposed in \cite{Seiberg:1999vs} we call $(g,b)$ the closed string data and $(G,\theta)$ the open string data. 
  We would like to stress
 one evident, but nevertheless important point: Despite the fact that the actions (\ref{actiondefbos})
 and (\ref{firstorderact})
 are classically equivalent we need slightly different geometrical data to define them. For the second 
 order action (\ref{actiondefbos}) we need $({\cal M}, g, H)$ while for the first order action (\ref{firstorderact})
 $({\cal M}, g, b , H)$. If $db=0$ then all first order actions with different $b$ are equivalent to the same
 second order action.  
 Hence two first order actions with $E^{\mu\nu}$ and $\tilde{E}^{\mu\nu}$ are physically equivalent if 
 either $E$ and $\tilde{E}$ are related by diffeomorphism or by a shift of the closed form (b-transform), 
 namely $(E_{\mu\nu}-\tilde{E}_{\mu\nu}) \in \Omega^2_{closed}({\cal M})$.   
 The symmetry group relating the different (but physically equivalent) first order actions  is 
 the semidirect product of $Diff({\cal M})$ and $\Omega^2_{closed}({\cal M})$.
 This observation will play an important role in further discussion\footnote{ In this context
 we have a comment which is not directly relevant to 
 the subject of this paper. Considering the properties of the first order 
 action we could define string theory in the following fashion: Choose 
 an open cover $\{ U_\alpha \}$ of a manifold ${\cal M}$. For each chart $U_\alpha$  
 define the first order action $S_\alpha$ using $E_\alpha$ and on the intersection
 $U_\alpha \cap  U_\beta$ glue the $E$'s using the semidirect product 
 of $Diff({\cal M})$ and $\Omega^2_{closed}({\cal M})$. Now $(G_\alpha, \theta_\alpha)$ 
 are not tensors in the usual sense anymore since we glue them on $U_\alpha \cap  U_\beta$
  using not only $Diff({\cal M})$. However  this ``exotic'' 
 prescription does not change the physics.
 This remark is related to the discussion of non-geometrical string theories in \cite{hmw, Flournoy:2004vn}.
 We hope to discuss these issues in detail elsewhere.}.   

Another interesting property is that the action (\ref{firstorderact}) includes the known 
 two-dimensional topological field theories as degenerate limits. Namely if $G=0$ and $d(B-b)=0$ then the action 
 (\ref{firstorderact}) corresponds to the Poisson sigma model introduced in \cite{Ikeda:1993fh, Schaller:1994es}, 
 provided that $\theta$ is a Poisson tensor. In the case $G=0$ and $d(B-b)\neq 0$ the model can be related to 
 more general type of topological theory, the WZ-Poisson sigma model \cite{Klimcik:2001vg}, assuming some
  specific differential condition between $\theta$ and $d(B-b)$. Presumably these topological models may arise
 as a result of a decoupling limit in string theory. Although the Poisson and WZ-Poisson sigma models are not 
 the main subject of this paper many results we present will be applicable to these models as well. 
    
 The main goal of this paper is to study the extended supersymmetry of the
first order action (\ref{firstorderact}). For technical reasons related to
supersymmetry it is convenient to switch to light-cone coordinates.
 Using $(1,0)$ superfields the $N=(1,0)$ supersymmetric
version of (\ref{firstorderact}) is
\beq
 S= i \int d^2\sigma\,d\theta\,\, \left ( D_+ \Phi^\mu S_{=\mu} - S_{+\mu}
 \d_= \Phi^\mu - S_{+\mu} S_{=\nu} E^{\mu\nu}  + D_+ \Phi^\mu \d_= \Phi^\nu (B-b)_{\mu\nu}
 \right )
\eeq{newact123AAA}
Throughout the paper we use $(\+, =)$ as worldsheet indices and $(+,-)$ as two-dimensional 
 spinor indices. We use $(1,0)$ superspace where with a spinor coordinate $\theta$. The
 covariant derivative $D_+$ and supersymmetry generator $Q_+$ satisfy
\beq
 D^2_+= i\d_\+,\,\,\,\,\,\,\,\,\,\,\,Q_+ =  iD_+ + 2\theta \d_\+
\eeq{nonsusy10}
 where $\d_{\pp} = \d_0 \pm \d_1$. In terms of the covariant derivatives, a supersymmetry
 transformation of a superfield $\Phi$ is given by
\beq 
 \delta_m \Phi = i\epsilon_m^+ Q_+ \Phi,\,\,\,\,\,\,\,\,\,\,\,\,
 \delta_m S_+ = i\epsilon_m^+ Q_+ S_+,\,\,\,\,\,\,\,\,\,\,\,\,
 \delta_m S_= = i\epsilon_m^+ Q_+ S_= .
\eeq{susymanif10}
 
In terms of $(1,1)$ superfields, the $N=(1,1)$ first order action is given by 
\beq
 S=  \int d^2\sigma\,d^2\theta\,\, \left ( D_+ \Phi^\mu S_{-\mu} - S_{+\mu}
 D_- \Phi^\mu - S_{+\mu} S_{-\nu} E^{\mu\nu}   + D_+ \Phi^\mu D_- \Phi^\nu (B-b)_{\mu\nu}\right ),
\eeq{newact123BBB}
 where we use the standard notation (see Appendix A in \cite{Lindstrom:2002jb}).
 In what follows we focus on $N=(1,0)$ models. We would like to understand 
 under which assumptions $N=(1,0)$ models admit $N=(2,0)$. However our results 
  may be straightforwardly generalized to the extension of $N=(1,1)$ to 
 $N=(2,1)$ susy.

\section{Generalized complex geometry}
\label{math}

In this Section we review some basic notions and fix notations. Namely we collect general facts  concerning 
 the generalized complex structure, see \cite{Hitchin} and \cite{Gualtieri} for further details. 
 Also we work out the coordinate 
 form of the integrability conditions for the generalized complex structure.

 Let us start by recalling the definition of the standard complex structure on a manifold ${\cal M}$ ($\dim {\cal M}=d$).
  An almost complex structure 
 is defined as a linear map on the tangent bundle $J : T \rightarrow T$ such that $J^2= -1_d$. This allows 
 the definition of projectors on $T$, 
\beq
\pi_\pm = \frac{1}{2}(1_d \pm i J).
\eeq{definbrale}
 An almost complex structure is called integrable if the projectors
  $\pi_\pm$ define integrable distributions on $T$, namely if
\beq
 \pi_\mp [\pi_\pm X, \pi_\pm Y] =0
\eeq{ingerabusualcoml}
 for any $X, Y \in T$ where $[\,\,,\,\,]$ is a standard Lie bracket on $T$.  

 A generalization of the notion of complex structure has been proposed by Hitchin \cite{Hitchin}. In Hitchin's construction 
  $T$ is replaced by $T\oplus T^*$ and the Lie bracket is replaced
 by the appropriate bracket on $T\oplus T^*$, the so called Courant bracket. Thus a generalized complex structure  is an almost 
 complex structure ${\cal J}$ on $T\oplus T^*$ whose $+i$-eigenbundle is Courant involutive.
This definition is the complex analog of a Dirac structure, a concept unifying Poisson and symplectic geometry,
  introduced by Courant and Weinstein \cite{courant,weinstein}. A detailed study of
 generalized complex geometry
 can be found in Gualtieri's thesis \cite{Gualtieri}. 

 Now let us give detailed definitions. On $T\oplus T^*$ there is a natural indefinite metric defined by 
 $(X+\xi, X+\xi ) =  i_X \xi$.
In the coordinate basis $(\d_\mu, dx^\mu)$ we can write this metric as follows
\beq
 {\cal I} = \left ( \begin{array}{ll}
                              0 & 1_d \\
                              1_d & 0 
                        \end{array} \right )
\eeq{definindefmetric}
 A generalized almost complex structure is a map ${\cal J}: T \oplus T^* \rightarrow T \oplus T^*$
 such that ${\cal J}^2 = -1_{2d}$ and that ${\cal I}$ is hermitian with respect to ${\cal J}$, ${\cal J}^t {\cal I}
 {\cal J} ={\cal I}$. On $T\oplus T^*$ there is a Courant bracket which is defined as follows
\beq
 [X +\xi, Y +\eta ]_c = [X, Y] + {\cal L}_X \eta - {\cal L}_Y \xi -\frac{1}{2} d(i_X \eta - i_Y\xi).
\eeq{defCourbrak}
 This bracket is skew-symmetric but in general does not satisfy the Jacobi identity. 
 However if there is a subbundle $L \subset T\oplus T^*$ which is involutive (closed under the Courant bracket) 
 and isotropic with respect to ${\cal I}$ then the Courant bracket on the sections of $L$ does satisfy 
 the Jacobi identity. This is a reason for imposing hermiticity of ${\cal I}$ with respect to ${\cal J}$.
 One important feature  of the Courant bracket is that, unlike the Lie bracket, this bracket has a nontrivial 
 automorphism defined by a closed two-form $b$,
\beq
 e^b(X + \xi)= X+\xi + i_X b.
\eeq{definaurom}
 such that 
\beq
 [e^b(X +\xi), e^b(Y +\eta) ]_c = e^b[X +\xi, Y +\eta ]_c .
\eeq{autmocourbracket} 
 We can construct the projectors on $T \oplus T^*$
\beq
 \Pi_{\pm} = \frac{1}{2} ( I \pm i {\cal J})\ ;
\eeq{defproj}
 the almost generalized complex structure ${\cal J}$ is integrable if 
\beq
  \Pi_{\mp} [\Pi_{\pm}(X+\xi), \Pi_{\pm}(Y+\eta)]_c = 0,
\eeq{inegrablproj} 
 for any $(X+\xi), (Y+\eta) \in T \oplus T^*$.
 This is equivalent to the single statement
\beq
 [X+\xi, Y+\eta]_c - [{\cal J}(X+\xi), {\cal J}(Y+\eta)]_c + {\cal J} [{\cal J}(X+\xi), Y+\eta]_c
 + {\cal J}[X+\xi, {\cal J}(Y+\eta)]_c =0
\eeq{coranaNEt}
 which resembles the definition of the Nijenhuis tensor. 

 To relate to the construction to the physical models we have to
 reexpress the above definitions in coordinate form. 
The map ${\cal J}$ can be written in the form  
\beq
 {\cal J} = \left ( \begin{array}{ll}
                      J & P \\
                      L & K 
                    \end{array} \right ),
\eeq{defgencomphne1}
 where $J: T{\cal M} \rightarrow T{\cal M}$,
 $P :T^*{\cal M} \rightarrow T{\cal M}$, $L: T{\cal M} \rightarrow T^*{\cal M}$ and $K: T^*{\cal M} \rightarrow 
 T^*{\cal M}$ and hence they correspond to the tensor fields, $J^\mu_{\,\,\,\nu}$, $L_{\mu\nu}$, $P^{\mu\nu}$
 and $K_\mu^{\,\,\,\nu}$. Then the condition ${\cal J}^2 = - 1_{2d}$ becomes 
\ber
\label{AA1}&&J^\mu_{\,\,\,\nu} J^\nu_{\,\,\,\lambda} + P^{\mu\nu} L_{\nu\lambda} = -\delta^\mu_{\,\,\,\lambda},\\
\label{AA2}&&J^\mu_{\,\,\,\nu} P^{\nu\lambda} + P^{\mu\nu}K_\nu^{\,\,\,\lambda} =0,\\
\label{AA3}&&K_\mu^{\,\,\,\nu} K_\nu^{\,\,\,\lambda} + L_{\mu\nu} P^{\nu\lambda} =  -\delta^\mu_{\,\,\,\lambda},\\
\label{AA4}&&K_{\mu}^{\,\,\,\nu} L_{\nu\lambda} + L_{\mu\nu} J^\nu_{\,\,\,\lambda} =0.
\eer{alegj2-1}
The hermiticity of ${\cal I}$ with respect to ${\cal J}$ translates into the following conditions
\beq
 J^\mu_{\,\,\,\nu} + K_{\mu}^{\,\,\,\nu} =0,\,\,\,\,\,\,\,\,\,\,\,
 P^{\mu\nu} = - P^{\nu\mu},\,\,\,\,\,\,\,\,\,\,\,
L_{\mu\nu}=-L_{\nu\mu}
\eeq{hermiticitycond}
 In local coordinates the integrability condition (\ref{coranaNEt}) is equivalent to the 
 following four conditions 
\ber
\label{alg1}&& J^\nu_{\,\,\,[\lambda} J^\mu_{\,\,\,\rho],\nu} + J^\mu_{\,\,\,\nu} J^\nu_{\,\,\,[\lambda,\rho]} 
 + P^{\mu\nu}L_{[\lambda\rho,\nu]}=0 \\
\label{alg2} &&P^{[\mu|\nu} P^{|\lambda\rho]}_{\,\,\,\,\,\,,\nu} = 0\\
 \label{alg3}&& J^\mu_{\,\,\,\nu,\rho} P^{\rho\lambda} + P^{\rho\lambda}_{\,\,\,\,\,\,,\nu} J^\mu_{\,\,\,\rho} - 
 J^{\lambda}_{\,\,\,\rho,\nu} P^{\mu\rho} + J^{\lambda}_{\,\,\,\nu,\rho} P^{\mu\rho} 
- P^{\mu\lambda}_{\,\,\,\,\,\,,\rho} J^\rho_{\,\,\,\nu}=0\\
\label{alg4} && J^\lambda_{\,\,\,\nu} L_{[\lambda\rho,\gamma]} +
  L_{\nu\lambda} J^\lambda_{\,\,\,[\gamma,\rho]} +J^\lambda_{\,\,\,\rho}
 L_{\gamma\nu,\lambda} +J^\lambda_{\,\,\,\gamma} L_{\nu\rho,\lambda} + L_{\lambda\rho} J^\lambda_{\,\,\,\gamma,\nu}
 + J^\lambda_{\,\,\,\rho} L_{\lambda\gamma,\nu}=0
\eer{inegrabcondalg}
 To summarize, the generalized complex structure ${\cal J}$ is defined by three tensor fields $J^\mu_{\,\,\,\nu}$, 
 $L_{\mu\nu}$ and $P^{\mu\nu}$ which satisfy the algebraic conditions (\ref{AA1})-(\ref{hermiticitycond})
  and the differential conditions 
 (\ref{alg1})-(\ref{alg4}).
 
The usual complex structure $J$ is embedded in the notion of generalized complex structure
\beq
 {\cal J} = \left ( \begin{array}{ll}
                    J & \,\,\,\,0 \\
                    0 & - J^t
           \end{array} \right ).
\eeq{examplecomlps}
 One can check that all properties (\ref{AA1})-(\ref{inegrabcondalg}) are satisfied provided that $J$ is
  a complex structure. 
 Also, a symplectic structure is an example of a generalized complex structure
\beq
{\cal J} = \left ( \begin{array}{ll}
                    0 & -\omega^{-1} \\
                    \omega & \,\,\,\,\, 0
           \end{array} \right )
\eeq{examlsymlpstr}
 where $\omega$ is an ordinary symplectic structure ($d\omega
 =0$). More exotic examples exist and are given by manifolds, 
  that  do not admit any known complex or symplectic structure, but do
 admit a generalized complex structure \cite{Gualtieri,gilgualtieri}.

 Consider a generalized complex structure ${\cal J}$; a new generalized complex structure 
 can be generated by 
\beq
 {\cal J}_b = \left ( \begin{array}{ll}
                    1 & 0 \\
                    b & 1
              \end{array} \right ) {\cal J} \left (\begin{array}{ll}
                    \,\,\,\,1 & 0 \\
                    -b & 1
              \end{array} \right )
\eeq{btrabsofrmb}
 if $b \in \Omega^2_{closed}({\cal M})$. The structure ${\cal J}_b$ is integrable due to the fact that 
 the transformation (\ref{definaurom}) is an automorphism of the Courant bracket.
The transformation (\ref{btrabsofrmb}) is called a $b$-transform and later we will 
 see that this is related to the $b$-transform for the first order actions discussed in the previous Section. 
 
The key feature of a complex manifold is that  is locally equivalent to $C^{k}$ via a diffeomorphism.
 For symplectic manifolds the Darboux theorem states that a symplectic structure is locally equivalent, via 
 diffeomorphism, to the standard symplectic structure $(R^{2k}, \omega)$, where
\beq
 \omega = dx_1 \wedge dx_2 + ... + dx_{2k-1} \wedge dx_{2k}.
\eeq{definsypmls}
 For  generalized complex manifolds there exists a generalized Darboux theorem \cite{Gualtieri}, 
 which states that in a neighborhood of a regular point\footnote{$P$ is a Poisson structure and it will 
 define a symplectic foliation. The point is called regular if $P$ has constant rank in a neighborhood.}
 a generalized complex structure on a manifold ${\cal M}$ is locally equivalent 
 via a diffeomorphism and a $b$-transform (see (\ref{btrabsofrmb})), to the product of an open set in $C^k$
 and an open set in the standard symplectic space $(R^{d-2k}, \omega)$.

 The Courant bracket on $T\oplus T^*$ can be twisted by a closed three form $H$. Namely given a closed
 three form $H$ one can define another bracket on $T\oplus T^*$ by
\beq
 [X +\xi, Y +\eta ]_H= [X +\xi, Y +\eta ]_c + i_X i_Y H.
\eeq{deftwCorbr} 
 This bracket has similar properties to the Courant bracket. 
 Again if a subbundle $L \subset T\oplus T^*$ is closed under the twisted Courant bracket 
 and isotropic with respect to ${\cal I}$, then the Courant bracket on the sections of $L$ does satisfy 
 the Jacobi identity.
Thus in the integrability condition (\ref{coranaNEt}) the Courant 
 bracket $[\,\,,\,\,]_c$ can be replaced by the new twisted
 Courant bracket $[\,\,,\,\,]_H$. 
 In local coordinates the new integrability condition is equivalent to four expressions
\ber
\label{alg1WZ} && J^\nu_{\,\,\,[\lambda} J^\mu_{\,\,\,\rho],\nu} + J^\mu_{\,\,\,\nu} J^\nu_{\,\,\,[\lambda,\rho]} 
 + P^{\mu\nu}(L_{[\lambda\rho,\nu]} + J^\sigma_{\,\,\,[\lambda}H_{\rho]\sigma\nu}  )=0 \\
\label{alg2WZ} &&P^{[\mu|\nu} P^{|\lambda\rho]}_{\,\,\,\,\,\,,\nu} = 0\\
 \label{alg3WZ} && J^\mu_{\,\,\,\nu,\rho} P^{\rho\lambda} + P^{\rho\lambda}_{\,\,\,\,\,\,,\nu} J^\mu_{\,\,\,\rho} - 
 J^{\lambda}_{\,\,\,\rho,\nu} P^{\mu\rho} + J^{\lambda}_{\,\,\,\nu,\rho} P^{\mu\rho} 
- P^{\mu\lambda}_{\,\,\,\,\,\,,\rho} J^\rho_{\,\,\,\nu}-P^{\lambda\sigma} P^{\mu\rho} H_{\sigma\rho\nu}=0 \\
\nonumber && J^\lambda_{\,\,\,\nu} L_{[\lambda\rho,\gamma]} + L_{\nu\lambda} J^\lambda_{\,\,\,[\gamma,\rho]}
  +J^\lambda_{\,\,\,\rho}
 L_{\gamma\nu,\lambda} +J^\lambda_{\,\,\,\gamma} L_{\nu\rho,\lambda} + L_{\lambda\rho} J^\lambda_{\,\,\,\gamma,\nu}
 + J^\lambda_{\,\,\,\rho} L_{\lambda\gamma,\nu} +\\
\label{alg4WZ} && + H_{\rho\gamma\nu} - J^\lambda_{\,\,\,[\rho} J^\sigma_{\,\,\,\gamma}
 H_{\nu]\lambda\sigma}=0 
\eer{inegrabcondalgWZ}

\section{Topological model}
\label{sec:top}

In this Section we consider a toy topological model which will provide a ``physical'' derivation of
generalized complex geometry. Also it will lead to results which will be relevant for 
 the physical model (\ref{newact123}). 
 The model has the following action 
\beq
 S_{top} =\int d^2\sigma\,d\theta\,\,S_{+\mu} \d_= \Phi^\mu  
\eeq{newtopact12}
 which is part of the action (\ref{newact123}). 
 This is a topological system which describes the  holomorphic maps $\Phi :\Sigma \rightarrow {\cal M}$. 
 The model  is manifestly  $N=(1,0)$ supersymmetric and can be defined over any differential manifold ${\cal M}$.
 We would like to find the restrictions on ${\cal M}$ arising from the requirement that 
 the model admits $(2,0)$ supersymmetry.  

 We have to look for additional (non manifest) supersymmetry transformations. 
  The general transformations of $S_+$ and 
 $\Phi$ are given by the following expressions
\beq
 \delta (\epsilon) \Phi^\mu = \epsilon^+ D_+\Phi^\nu J^\mu_{\,\,\,\nu} - \epsilon^+ S_{+\nu} P^{\mu\nu},
\eeq{neesphi}
\ber
\nonumber  \delta (\epsilon) S_{+\mu}= i \epsilon^+ \d_{\+} \Phi^\nu L_{\mu\nu} - \epsilon^+
 D_+ S_{+\nu} K_\mu^{\,\,\,\nu} +
  \epsilon^+ S_{+\nu} S_{+\rho} N_{\mu}^{\,\,\,\nu\rho} +\\
+ \epsilon^+ D_+ \Phi^\nu D_+\Phi^\rho  M_{\mu\nu\rho}
 + \epsilon^+ D_+ \Phi^\rho S_{+\nu} Q_{\mu\rho}^{\,\,\,\,\,\,\nu}
\eer{newsSplus}
 Classically the Ansatz (\ref{neesphi}) and (\ref{newsSplus}) is unique on
  dimensional grounds and by Lorentz covariance \cite{Lindstrom:2004eh}. 
 This Ansatz involves seven different tensors on ${\cal M}$.  
  We have to require the standard $N=(2,0)$ supersymmetry algebra, i.e. the manifest and non-manifest
  supersymmetry transformations commute and the nonmanifest supersymmetry transformations satisfy   
 the following conditions
\beq
 [\delta(\epsilon_2), \delta(\epsilon_1)]\Phi^\mu = 2i \epsilon^+_1 \epsilon_2^+ \d_{\+} \Phi^\mu,\,\,\,\,\,\,\,\,\,\,\,\,\,
 [\delta(\epsilon_2), \delta(\epsilon_1)]S_{+\mu}=  2i \epsilon^+_1 \epsilon_2^+ \d_{\+} S_{+\mu}.
\eeq{summartwhat}
 Since the nonmanifest transformations are written in $(1,0)$ superfield then the first requirement is 
 automatically satisfied. Next we have to calculate the commutator of two nonmanifest supersymmetry transformations. 
 The result of the calculation is given in Appendix. Imposing the condition (\ref{summartwhat}) implies
 four algebraic and eleven differential conditions on the seven tensors introduced in (\ref{neesphi}) 
 and (\ref{newsSplus}). This fact alone shows how the problem of extended supersymmetry becomes
 involved when extra fields are introduced.   

 Before analyzing the algebra in detail it is useful to look at the 
invariance of the action.
The action (\ref{newtopact12}) is invariant under (\ref{neesphi}) and (\ref{newsSplus}) if 
 the following algebraic conditions 
\beq
 J^\mu_{\,\,\,\nu} + K_\nu^{\,\,\,\mu} =0,\,\,\,\,\,\,\,\,\,\,\,
 L_{\mu\nu} = - L_{\nu\mu},\,\,\,\,\,\,\,\,\,\,\,
 P^{\mu\nu} = - P^{\nu\mu}
\eeq{algebrcontop}
 as well as the differential conditions
\beq
 \frac{1}{2} P^{\mu\nu}_{\,\,\,\,\,\,,\rho} = -N_\rho^{\,\,\,\mu\nu},\,\,\,\,\,\,\,\,\,
 J^\mu_{\,\,\,[\nu,\rho]} = Q_{\nu\rho}^{\,\,\,\,\,\,\mu},\,\,\,\,\,\,\,\,\,
 \frac{1}{2} L_{[\mu\nu,\rho]}= M_{\rho\nu\mu}
\eeq{diftopconac}
 are satisfied. The differential conditions (\ref{diftopconac}) allow us to express all three index tensors
 in terms of appropriate derivatives of two index tensors $J$, $P$, $L$ and $K$. 
 These two index tensors can be combined as a single object
\beq
 {\cal J} = \left ( \begin{array}{ll}
                      J & P \\
                      L & K 
                    \end{array} \right ),
\eeq{defgencomph}
 where ${\cal J}: T \oplus T^* \rightarrow T \oplus T^*$.
 It is easy to see that the algebraic part of the 
supersymmetry algebra (the part of (\ref{alfphi123},\ref{alpsplus12}) 
which does not involve derivatives nor three--index tensors)
can be written as a single equation, namely that  
${\cal J}^2 = - 1_{2d}$. Passing then to the action, 
the algebraic condition (\ref{algebrcontop}) is equivalent to a hermiticity of 
 ${\cal I}$ with respect to ${\cal J}$ (i.e., the natural pairing on 
$T\oplus T^*$, see 
 previous Section). Therefore ${\cal J}$ is an almost generalized complex structure.
 Finally we have to analyze the eleven differential conditions coming from the algebra using
  (\ref{diftopconac}). Using the results from the previous Section, we see that the three differential conditions 
 arising from (\ref{alfphi123}) are the same as the conditions (\ref{alg1})-(\ref{alg3}).  
  The second line in (\ref{alpsplus12}) is equivalent to the condition (\ref{alg4}).  
Surprisingly the remaining differential conditions in (\ref{alpsplus12})
 are automatically satisfied provided that (\ref{alg1})-(\ref{alg4}) hold and
 ${\cal J}$ is a almost generalized complex structure. Therefore we have proved that the differential conditions 
 that come from the supersymmetry algebra are equivalent to integrability of ${\cal J}$ with respect to the Courant 
 bracket. 

To summarize the topological model (\ref{newtopact12}) admits $(2,0)$ supersymmetry if and only 
 if  manifold ${\cal M}$ is generalized complex manifold. 

 As we briefly mentioned in the previous Section, a generalized complex manifold is equivalent locally, via diffeomorphism 
and $b$-transform, to a product of a symplectic and a complex manifolds. If we choose the Darboux coordinates (label $n$)
 along the symplectic leaf and the standard complex coordinates (label $i$, $\bar{i}$) transverse to the leaf
 the supersymmetry transformations (\ref{neesphi}) and (\ref{newsSplus}) is simplified drastically  and have the 
 following form
\beq
 \delta \Phi^i = i \epsilon^+ D_+ \Phi^i,\,\,\,\,\,\,\,\,\,\,\,\,\,\,\,\delta \Phi^{\bar{i}} =
 - i\epsilon^+ D_+\Phi^{\bar{i}}
\eeq{complecoor1}
\beq
 \delta S_{+i} = i \epsilon^+ D_+ S_{+i},\,\,\,\,\,\,\,\,\,\,\,\,\,\,\,\delta S_{+\bar{i}} = - i\epsilon^+ D_+ S_{+{\bar{i}}}
\eeq{complexcoor2}
\beq
 \delta \Phi^n = -\epsilon^+ S_{+(n+1)},\,\,\,\,\,\,\,\,\,\,\,\,\,\,\,\delta S_{+(n+1)} = - i\epsilon^+ \d_\+ \Phi^n
\eeq{sympecoord1}
\beq
\delta \Phi^{n+1} = \epsilon^+ S_{+n},\,\,\,\,\,\,\,\,\,\,\,\,\,\,\,
 \delta S_{+n}  = i \epsilon^+ \d_\+ \Phi^{n+1}
\eeq{symplscoord2}

\section{Topological model with WZ term}

In the previous Section we presented the topological model for which the extended supersymmetry 
 is related to the generalized complex structure with integrability defined with the respect 
 to the Courant bracket. The  natural question is now the following: if in the integrability condition the Courant bracket is 
 replaced by the twisted Courant bracket, 
can we then construct a model which incorporates
 twisted integrability? This is in fact possible and the solution is related to the WZ term.

 We consider the topological model with an additional term
\beq
 S_{top} = \int d^2\sigma\,d\theta\,\,S_{+\mu} \d_= \Phi^\mu  - \frac{1}{2} \int d^2\sigma\,d\theta\,\,D_+ \Phi^\mu
 \d_= \Phi^\nu B_{\mu\nu}
\eeq{acttopwzter}
 The last term is a WZ term and it depends only on a closed three-form $H$
\beq
 H_{\mu\nu\lambda} = \frac{1}{2}( B_{\mu\nu,\lambda} + B_{\lambda\mu,\nu} + B_{\nu\lambda,\mu}) 
\eeq{defindH}
if the world-sheet does not have a boundary. The model (\ref{acttopwzter}) has $N=(1,0)$ supersymmetry
 and can be  defined over any differential manifold ${\cal M}$ equipped with a closed three-form $H$. 
 The Ansatz for the nonmanifest supersymmetry transformations is given by the same expressions
 as before, (\ref{neesphi}) and (\ref{newsSplus}). The off-shell supersymmetry  algebra is exactly the 
 same, (\ref{summartwhat}).

 The main difference comes from the action. Namely invariance of the new action (\ref{acttopwzter})
 under the transformations (\ref{neesphi}) and (\ref{newsSplus})
  leads to new relations between the three and two index
 tensors in the supersymmetry transformations. 
The action (\ref{acttopwzter}) is invariant under (\ref{neesphi}) and (\ref{newsSplus}) if 
 the following algebraic conditions are satisfied
\beq
 J^\mu_{\,\,\,\nu} + K_\nu^{\,\,\,\mu} =0,\,\,\,\,\,\,\,\,\,\,\,
 L_{\mu\nu} = - L_{\nu\mu},\,\,\,\,\,\,\,\,\,\,\,
 P^{\mu\nu} = - P^{\nu\mu}
\eeq{algebrcontopWZ}
 as well as the differential conditions
\beq
 \frac{1}{2} P^{\mu\nu}_{\,\,\,\,\,\,,\rho} = -N_\rho^{\,\,\,\mu\nu},\,\,\,\,\,\,\,\,\,
 J^\mu_{\,\,\,[\nu,\rho]} + P^{\mu\lambda} H_{\lambda\nu\rho} = Q_{\nu\rho}^{\,\,\,\,\,\,\mu},\,\,\,\,\,\,\,\,\,
 \frac{1}{2} L_{[\mu\nu,\rho]} + \frac{1}{2} J^\lambda_{\,\,\,[\mu}H_{\nu]\lambda\rho}= M_{\rho\nu\mu}.
\eeq{diftopconacWZ}
  The algebraic part of all conditions remains the same as in the previous Section and therefore the 
 two-index tensors can be combined in a single object ${\cal J}$ which is an almost generalized 
 complex structure. However the differential conditions will change. Using (\ref{diftopconacWZ})
 we have to require that the expressions (\ref{alfphi123}) and (\ref{alpsplus12}) reproduce
  the supersymmetry algebra (\ref{summartwhat}).
 Using the results from  Section \ref{math} we see that the three differential conditions 
 arising from (\ref{alfphi123}) are the same as conditions (\ref{alg1WZ})-(\ref{alg3WZ}).  
  The second line in (\ref{alpsplus12}) is equivalent to the condition (\ref{alg4WZ}).  
 As before the remaining differential conditions in (\ref{alpsplus12})
 are automatically satisfied provided that (\ref{alg1WZ})-(\ref{alg4WZ}) hold and that
 ${\cal J}$ is a almost generalized complex structure. Therefore we have proved that the differential conditions 
 coming from the supersymmetry algebra are equivalent to integrability of ${\cal J}$ with respect to the {\em twisted} 
 Courant 
 bracket.

\section{Sigma model}

 Now we turn to the ``real'' sigma model. For the sake of clarity, let
 us assume that the WZ term is absent in the action. Thus the second order $N=(1,0)$ 
 action is given by
\beq
 S= -i \int d^2\sigma\,d \theta\,\,D_+\Phi^\mu \d_= \Phi^\nu E_{\mu\nu}(\Phi)  .
\eeq{noract10}
 This action has the following first order form
\beq
 S= i \int d^2\sigma\,d\theta\,\, \left ( D_+ \Phi^\mu S_{=\mu} - S_{+\mu}
 \d_= \Phi^\mu - S_{+\mu} S_{=\nu} E^{\mu\nu} \right ).
\eeq{newact123}
 Again, we would like to study under which conditions on the geometry of ${\cal M}$
 the model (\ref{newact123}) admits $(2,0)$ supersymmetry. 

 We start by giving the most general Ansatz for the nonmanifest supersymmetry transformations.
 We already gave the most general Ansatz for the transformations of $S^\mu_+$ and $\Phi^\mu$, 
see (\ref{neesphi}) and (\ref{newsSplus}). For $S_=$ we can write the following most general 
 classical Ansatz for the transformations \cite{Lindstrom:2004eh}
\ber
 \nonumber  \delta (\epsilon) S_{=\mu} = \epsilon^+ D_+ S_{=\nu} R_{\mu}^{\,\,\,\nu} + \epsilon^+
 \d_= S_{+\nu}  Z_{\mu}^{\,\,\,\nu} + \epsilon^+ D_+ \d_= \Phi^\nu T_{\mu\nu} + \\
+ \epsilon^+ S_{+\rho} \d_= \Phi^\nu U_{\mu\nu}^{\,\,\,\,\,\,\rho} + \epsilon^+ D_+ \Phi^\nu S_{=\rho}
 V_{\mu\nu}^{\,\,\,\,\,\,\rho} + \epsilon^+ D_+ \Phi^\nu \d_= \Phi^\rho X_{\mu\nu\rho} +
 \epsilon^+ S_{+\nu} S_{=\rho} Y_{\mu}^{\,\,\,\nu\rho} .
\eer{newsSminusminus}
 Thus altogether the supersymmetry transformations contain 14 different tensors. 
 The commutators of non-manifest supersymmetry transformations are given in Appendix. We have to 
 require that (\ref{alfphi123}) and (\ref{alpsplus12}) reduces to (\ref{summartwhat}) (off-shell supersymmetry algebra) and 
 that (\ref{BIGSM}) reduces to
\beq
 [\delta(\epsilon_2), \delta(\epsilon_1)]S_{=\mu}=  2i \epsilon^+_1 \epsilon_2^+ \d_{\+} S_{=\mu}.
\eeq{trasnS--need}

 The action (\ref{newact123}) is invariant under the transformations (\ref{neesphi}), (\ref{newsSplus})
  and (\ref{newsSminusminus})
 if the following algebraic conditions are satisfied
\ber
 \label{SM1} J^\nu_{\,\,\mu} + L_{\rho\mu} E^{\rho\nu} + R_\mu^{\,\,\,\nu} = 0 \\
\label{SM2} P^{\nu\mu} + E^{\rho\nu} K_\rho^{\,\,\,\mu} + E^{\mu\rho} R_{\rho}^{\,\,\,\nu}=0 \\
\label{SM3}L_{(\nu\mu)}+T_{(\mu\nu)}=0 \\
\label{SM4} Z_\rho^{\,\,\,(\mu}E^{\nu)\rho}-P^{(\mu\nu)} = 0 \\
\label{SM5}  J^\mu_{\,\,\nu} + T_{\rho\nu} E^{\mu\rho} -Z_\nu^{\,\,\,\mu} + K_\nu^{\,\,\,\mu} =0 
\eer{Paleginvact20}
 as well as the following differential conditions
\ber
 \label{SMDC1} J^\mu_{\ \, [\nu,\rho]} -V_{[\nu\rho]}^{\,\,\,\,\,\,\mu} - 
 M_{\lambda[\nu\rho]} E^{\lambda\mu} + R_{[\nu\,\,\,,\rho]}^{\,\,\,\mu}=0 \\
 \label{SMDC2}   P^{\mu\lambda}_{\ \ ,\rho}- E^{\lambda\nu} V_{\nu\rho}^{\ \, \mu}
 - Y_\rho^{\,\,\,\lambda\mu} -
 Q_{\nu\rho}^{\,\,\,\,\,\,\lambda} E^{\nu\mu} + (E^{\lambda\nu}R_\nu^{\,\,\,\mu})_{,\rho} - E^{\lambda\mu}_{\,\,\,\,\,\,,\nu}
 J^\nu_{\,\,\,\rho} = 0\\
\label{SMDC3} - U_{\lambda\mu}^{\,\,\,\,\,\,\rho} -
E^{\rho\nu} X_{\nu\lambda\mu} - Q_{\mu\lambda}^{\,\,\,\,\,\,\rho} - 
  J^\rho_{\,\,\,\lambda,\mu} + Z^\rho_{\,\,\,\lambda,\mu} - K^\rho_{\,\,\,\mu,\lambda}= 0 \\
\label{SMDC4} X_{[\mu\lambda]\rho} + M_{\rho[\mu\lambda]} - \frac12
T_{[\mu\lambda],\rho}+L_{\rho[\mu,\lambda]} + \frac12 L_{[\mu\lambda],\rho{}}= 0\\
\label{SMDC5}  \frac12 ( Z_\nu^{\ [\mu} E^{\rho]\nu})_{, \,\lambda} 
- E^{[\mu|\nu} U_{\nu\lambda}^{\,\,\,\,\,\,|\rho]} +
N_{\lambda}^{\,\,\,[\mu\rho]} +\frac12 P^{[\mu\rho]}_{\,\,\,\,\,\,\ ,\lambda}=0\\
\label{SMDC6} - Y_\nu^{\,\,\,[\lambda|\rho}E^{|\mu]\nu} + E^{\nu\rho} 
N_\nu^{\,\,\,[\mu\lambda]} + E^{[\mu|\rho}_{\,\,\,\,\,\,,\nu}
  P^{\nu|\lambda]} = 0
\eer{difconinvact20}
 Combining these conditions with the supersymmetry algebra we may analyze the solutions 
 of the problem. In particular we are interested in the geometrical interpretation of the solutions.
 We will see that to find a general solution is hard. This is partially due to absence of appropriate 
 mathematical notions. However we will present the solution related to the generalized complex structure 
 as defined by Hitchin \cite{Hitchin}. 
 Regarding more general solutions, we can offer only some 
speculations, presented in subsection \ref{sec:diffcond}.  

Before turning to a discussion of possible solutions, we caution the reader that the general Ansatz we have
 made for the second supersymmetry will have solutions that correspond to ``field equation''--type symmetries, 
as discussed in \cite{Lindstrom:2004eh}. E.g.,  any transformation of the form
\beq
\delta S_{+\mu}=\epsilon^+A_{\mu\nu}D_+F^\nu_+~, \qquad \delta S_{=\mu}=\epsilon^+D_+(A_{\nu\mu}F^\nu_=)~,
\eeq{}
 will be a ``trivial'' symmetry of the $(2,0)$ action (\ref{newact123}) if $F^\nu_+$ and $F^\nu_=$ are the $S_{=\mu}$ and
 $S_{+\mu}$ field equations, respectively. 

\subsection{Algebraic conditions}

In this section we will analyze the content of the algebraic conditions coming
from invariance of the action, (\ref{SM1}--\ref{SM5}), and 
from the algebraic part of the closure of the algebra, (\ref{alfphi123},
\ref{alpsplus12}, \ref{BIGSM}). For the topological model in section
\ref{sec:top}, we were able to reformulate all conditions in terms of an
almost complex structure ${\cal J}$. Here we will try to get as close as we
can to this doing the same for the sigma model, in particular we try to reexpress all 
conditions, now written in terms of $d\times d$
matrices, in terms of big $2d\times 2d$ matrices. The reason for this is 
 to make contact with
 the generalized structures. In the case at hand, the
geometry can even be analyzed in terms of the usual geometric structures on the
manifold (and not on $T \oplus T^*$), analogously to the case dubbed
``generalized K{\"a}hler structure'' in \cite{Gualtieri}. (We will find the algebraic
conditions of that case as an important particular case.) 

 We start by considering the conditions coming from 
the action.
For example, equations (\ref{SM1},\ref{SM2}) can be written more
elegantly as 
\begin{equation}
  \label{eq:vecstabilizer}
\left(\begin{array}{cc}
J^t& L^t\\ P^t& K^t
  \end{array}\right)
\left(
  \begin{array}{c}
E\\1
  \end{array}\right)
=
- \left(
  \begin{array}{c}
E\\1
  \end{array}\right)
E^{-1} R E\ .
\end{equation}
In what follows, we will refer to $d\times 2d$ matrices such as the one in 
(\ref{eq:vecstabilizer}) as ``vectors'', so that the equation itself we can 
be thought of as the vector $E \choose 1$ being stabilized by the matrix
${\cal J}^t$, with ``eigenvalue'' $(-E^{-1} R E)$.  If we define the
projective action of $GL(2d)$ on $d$-dimensional matrices as 
${{A \ B}\choose {C \ D}} \cdot E = (A E+ B) (CE+D)^{-1}$, it is easy to
eliminate the eigenvalue from (\ref{eq:vecstabilizer}), to find 
\begin{equation}
  \label{eq:stabilizer}
 {\cal J}^t\cdot E =E\ .
\end{equation}
In the old notation this equation reads $J^t E + L^t= E(P^t E + K^t)$, which
  means that ${\cal J}^t$ stabilizes  $E$  under the projective action.

Turning to equations (\ref{SM3}, \ref{SM4},
\ref{SM5}), we put them in the form 
\begin{equation}
  \label{eq:newherm}
\widetilde{\cal J}^t + {\cal I} \widetilde {\cal J } {\cal I}=0\ , \qquad   
 \widetilde {\cal J} \equiv  {\cal J} +
\left( \begin{array}{c} E^{-1} \\ 1 \end{array}\right) (T, -Z) 
= 
\left(\begin{array}{cc}
J+E^{-1}T & P-E^{-1} Z\\ L + T & K- Z  
\end{array}\right)\ .
\end{equation}
Here ${\cal I}$ is again the metric ${0\ 1}\choose {1\ 0}$  in
(\ref{definindefmetric}). As explained there, the usual hermiticity condition 
 for this metric reads ${\cal J}^t {\cal I}
 {\cal J} ={\cal I}$. Hence (\ref{eq:newherm}) is a hermiticity
 condition for $\widetilde {\cal J}$. We hasten to add that so far nothing
 says that this hermitian object squares to minus the identity, as was the
 case for ${\cal J}$ in the previous contexts. In fact, we shall see
 that in general this is  not
  the case. 

We now move to conditions coming from closure of the algebra. Fortunately, the
algebraic parts in (\ref{alfphi123},\ref{alpsplus12}) were already analyzed in 
section \ref{sec:top}. It is noticed there that they can be rewritten as the
condition ${\cal J}^2=-1_{2d}$. Condition (\ref{BIGSM}) is harder and requires
more care. Collecting the algebraic part gives the equations
\begin{equation}
  \label{eq:s=alg}
  R Z + Z K - TP =0\ , \quad RT-ZL+TJ=0\ , \quad R^2=-1\ .
\end{equation}
The first two of these read more compactly  
\begin{equation}
  \label{eq:compact}
  (T, -Z) 
\left(  \begin{array}{cc}
J&P\\L&K
  \end{array}\right)
= -R \ (T, -Z) \ .
\end{equation}
Again, these condition can be thought of as a stabilization. 
As for the third condition in (\ref{eq:s=alg}), we will show shortly that it
is implied by the other conditions we already have. 

(Before moving on, as a curiosity, we also notice that we can  
combine all  of (\ref{eq:s=alg}) with ${\cal J}^2=-1_{2d}$, to give
\begin{equation}
  \label{eq:curious}
\left(  \begin{array}{ccc}
R & T & -Z\\ 0 & J & P\\ 0 & L &K 
  \end{array}\right)^2 = -1_{3d}
\end{equation}
which thus summarizes all the algebraic equations from the algebra.)

We have now rewritten all conditions in ones that involve 
$2d\times 2d$ matrices. We use this 
to make contact with generalized structures. First of all, for the reader's
convenience we list the algebraic conditions we have found:
\begin{enumerate}
\item ${\cal J}^2=-1_{2d}$ (from closure of the algebra, 
(\ref{alfphi123},\ref{alpsplus12}));
\item ${\cal J}^t {E \choose 1} = -{E \choose 1} E^{-1} R E$ 
(\ref{eq:vecstabilizer});
\item ${\cal J}^t {T^t \choose -Z^t} = -{T^t\choose -Z^t} R^t$ (\ref{eq:compact});
\item $ \widetilde{\cal J}+ {\cal I} \widetilde {\cal J } {\cal I}=0$, 
where 
$\widetilde {\cal J} \equiv  {\cal J} +
{E^{-1} \choose 1}(T, -Z)$, eq. (\ref{eq:newherm}).
\end{enumerate}

Having a list of objects on $T \oplus T^*$ and their conditions, it 
 would seem natural at this point to ask to which subgroup
of SO$(d,d)$ they reduce. Unfortunately the conditions are not enough to 
determine a structure; there are many possible cases.  
This may be seen from the fact that conditions
2. and 3. may be more or less restrictive, depending on how 
many columns ${T^t \choose -Z^t}$ and ${E \choose 1}$ have in common. So they
can range from $d$ to $2d$ independent conditions. 

To see this more explicitly, it is useful to change to a basis in which ${\cal
  J}$ simplifies. That this may be possible is again suggested by conditions
  2. and 3. above: in the extreme case in which all columns of 
${T^t \choose -Z^t}$ and ${E \choose 1}$ are all independent, they 
can be regarded as a basis in which ${\cal J}$ is block--diagonal. Rather than
  doing this, we will display another change of basis, which does not rely on any
  assumption about the rank of ${{T^t \ E}  \choose {-Z^t \ 1}}$. 



The idea is to get another vector which is stabilized by ${\cal
  J}$, and which cannot have any column in common with one of those we already
  have, ${E \choose 1}$. The condition that this be stabilized,
(\ref{eq:vecstabilizer}), implies indeed that also an
orthogonal vector is stabilized: 
$$
(1, -E) {\cal J}^t \left( \begin{array}{c} E \\ 1 \end{array}\right) = 0
\ .
$$
This is seen to imply that 
\begin{equation}
  \label{eq:leftEstab}
(1, -E) {\cal J}^t=J_-^t(1,-E)  
\end{equation}
for some $J_-$. 

One might hope this
result, along with (\ref{eq:vecstabilizer}), can be used to produce a
block--diagonalizing change of basis. However, to do that we need both
right actions or both left actions. But if we transpose
(\ref{eq:leftEstab}), we get a statement on ${\cal J}$ and not ${\cal
  J}^t$. A way out of this situation would be to have a hermiticity
condition related to  ${\cal J}$; we do not have this, but the next best is
condition 4. above, (\ref{eq:newherm}).\footnote{Another possibility would
  have been that the hermitian object, ${\cal J} + X$, 
also squared to minus one. Unfortunately one finds 
$({\cal J}+ X)^2 = -1 + {\cal I} X^t {\cal I} X$, a relation similar to
U$(d)$ structures on manifolds of dimension higher than $d$.}
Defining 
$X= {E^{-1} \choose 1}(T, -Z)$, this gives us
$$
{\cal J}^t \left( \begin{array}{c} -E^t \\ 1 \end{array}\right) =
-\left( {\cal I} {\cal J} {\cal I} + X + {\cal I} X^t {\cal I} \right)
\left( \begin{array}{c} -E^t \\ 1 \end{array}\right) =
-\left( \left( \begin{array}{c} -E^t \\ 1 \end{array}\right) J_- +
\left( \begin{array}{c} 1 \\ E^{-1} \end{array}\right) 
(ZE^t+T)\right)\ .
$$
With this further computation, and using the action of ${\cal J}$ on
the other block--vector (\ref{eq:vecstabilizer}), we 
obtain
\begin{equation}
  \label{eq:triang}
  {\cal J}^t= -{\cal I} {\cal E} 
 \left( \begin{array}{cc}
E^{-1} R E & E^{-1} \theta \\ 0& J_-
  \end{array}\right)({\cal I} {\cal E})^{-1}\ , \qquad {\cal E}\equiv 
\left(\begin{array}{cc}
1&1 \\ E & -E^t
\end{array}\right)\ ,  \qquad \tau\equiv ZE^t+T\ .
\end{equation}
We have a basis in which ${\cal J}$ is block--triangular.
Although we have not used condition 3. yet, this form already shows that the
stabilizer depends on the off--diagonal block $\tau$. Rather than to attempt a
complete classification, we now show that the geometry can be described in terms
of tensors on the manifold (which is not always the case in generalized
complex geometry) and then return to the $T \oplus T^*$    point of view 
examining an important example.

The geometry can be analyzed in terms of tensors on the manifold for a
simple reason. 
It is immediate to notice that the condition ${\cal J}^2=-1$ implies that
\begin{equation}
  \label{eq:acs}
  R^2=-1\ , \qquad J_-^2=-1\ , \qquad R \tau = \tau J_-\ ;
\end{equation}
that is, $R$ and $J_-$ are two almost complex structures, and $\tau$ is an
intertwiner between them. (These facts could have been obtained
  without the change of basis; e.g., it is easy to show that
condition 2.~alone is enough to assume that the ``eigenvalue'' $R$ squares to
minus one, and similarly from (\ref{eq:leftEstab}) for $J_-$.) The fact that 
$R$ squares to minus one also came more directly from the action,
 (of (\ref{eq:s=alg})); here we showed that it is a consequence of the
other conditions 1.--4.~above. This is why it was not included in that
list. 

We still have one condition that we
have not used, 3.~in the list above, equation
(\ref{eq:compact}). The condition
is best analyzed after the ${\cal I E}$ change of basis. There, 
(\ref{eq:compact}) reads
\begin{equation}
  \label{eq:thz}
  \left( \begin{array}{cc}
E^{-1} R E & E^{-1} \tau \\ 0& J_-
  \end{array}\right)\left( \begin{array}{c} \zeta \\ -\frac12 g^{-1}\tau \end{array}\right) =
\left( \begin{array}{c} \zeta \\ -\frac12 g^{-1}\tau
  \end{array}\right)R^t \ ; 
\end{equation}
 $\zeta\equiv -\frac12 g^{-1}(T-ZE)^t$, and  $\tau$
 appears both in the matrix and in the vector, which makes the problem
 quadratic. Indeed, massaging the two components of these equations gives
 \begin{equation}
   \label{eq:tzherm}
   \tau (J_- - g^{-1}J_-^t g)=0\ , \qquad
   R(E\zeta)-(E\zeta)R^t=\frac12 \tau g^{-1} \tau^t\ .
 \end{equation}
These equations are modified (anti)--hermiticity properties on
the two almost complex structures $R$ and $J_-$. 

In summary, as seen from equations 
(\ref{eq:acs}) and (\ref{eq:tzherm}), there exists
 two almost complex structures, $R$ and $J_-$ on the manifold, with an intertwiner
between them, $\tau$; the two almost complex structures are antihermitian 
one on the image and one on the kernel of this intertwiner. Notice
that $J_-$ is equal to the almost complex structure of the
usual sigma model (\ref{noract10}) after integrating out the fields $S$ from
the first order action (\ref{newact123}).

\subsubsection{The hermitian case}

 We now  analyze an example, from both the $T\oplus
T^*$ and the $T$ perspectives. Above, the problematic point was that
  the object which squares to minus one, ${\cal J}$, and the
object which is hermitian, ${\cal J} + {E^{-1} \choose 1}(T, -Z)$, were not
the same. 

To overcome this, we assume in this subsection that 
${\cal J}^t + {\cal I} {\cal J} {\cal I} =0$ (i.e., ${\cal J} \in O(d,d)$).
Our previous formulae then reduce to the 
``generalized K{\"a}hler'' geometry of \cite{Gualtieri}, at least as far as algebraic
conditions are concerned.

 We start from the fact that ${\cal J}^t$ stabilizes $E=g+b$. Under the
 new hermiticity assumption, this is equivalent to the following 
statement:
\beq
 [{\cal J}, G] = 0\ .
\eeq{comgjetc}
 Here $G$ is a metric of signature $d,d$ defined as \cite{Gualtieri}
\beq
 G = \left ( \begin{array} {ll}
              -g^{-1} b & g^{-1} \\
               g - bg^{-1} b & b g^{-1}
               \end{array} \right ) = -1_{2d} + \left (\begin{array}{l} 
                                                 1 \\
                                                 E \end{array} \right ) g^{-1} (E^t\,\,\,1)
\eeq{defGualG}
with the property that $G^2=1_{2d}$ and $G^t {\cal I} G = {\cal I}$.
$G$ (or $E$) reduces the structure group on $T\oplus T^*$ 
to O$(d)\times$O$(d)$. ${\cal J}$ reduces the structure to U$(d/2,d/2)$. 
Together, 
and with the compatibility condition (\ref{comgjetc}) (or ${\cal J}\cdot E 
=E$), they reduce to U$(d/2)\times$U$(d/2)$. 

Equation (\ref{comgjetc}) can be shown formally from the stabilization
  condition, but is 
  particularly easy to see in the basis introduced above. From 
(\ref{eq:triang}) and (\ref{defGualG}), one gets
\begin{equation}
  \label{eq:commbasis}
  {\cal J}= {\cal E} \left( \begin{array}{cc}
E^{-1} R E & \\ 0& J_-
  \end{array}\right){\cal E}^{-1} \ , \qquad
G={\cal E} \left( \begin{array}{cc}
1 & \\ 0& -1
  \end{array}\right){\cal E}^{-1}\ , \qquad
I= {\cal E} \left( \begin{array}{cc}
g & \\ 0& -g
  \end{array}\right){\cal E}^{-1}
\end{equation}
 The first equation in (\ref{eq:commbasis}) 
is the same as equation (6.3) in \cite{Gualtieri}, after
redefining $J_+\equiv E^{-1} R E $. (There, the change of basis ${\cal E}$
has been factorized as ${{1 \ \ 0}\choose {b \ \ 1}} 
{{1 \ \ \ 1} \choose {g \ -g}}$.) As for the general case,
$J_\pm$ are almost complex structures. However, given the hermiticity
assumption, the form of 
the pairing ${\cal I}$ in (\ref{eq:commbasis}) also shows that these two
almost complex structures are both hermitian with respect to the metric $g$. 
This geometry is called (almost) bi--hermitian on the manifold. 

 We are not done yet, because imposing hermiticity does not set
$T$ and $Z$ to zero. What one gets is the remnant of (\ref{eq:newherm}), that
is, $X + {\cal I} X^t {\cal I}$. In components, this gives
$T^t = -T$ and $Z^t = E^{-1} T$. Using condition 3.~yields
 $RT^t -T R^t =0$, hence, $T$ is an intertwiner between $R$ and its 
transpose. Equivalently, we might want to define the matrix
\beq
 \hat{\cal J} = \left (\begin{array}{ll}
                       R^t & 0 \\
                       T & -R
                       \end{array} \right )
\eeq{defonnewalcs}
which is then an almost generalized complex structure.

It is also interesting to see what happens if we slightly relax the
initial condition. Looking at the triangular form
for ${\cal J}$, (\ref{eq:triang}), 
a natural condition  is $\tau=0$. ($\tau$ is only one
component of $X + {\cal I} X^t {\cal I}$, and thus this is weaker than the
hermiticity considered above.) 
In this case, we have a
condition similar to (\ref{comgjetc}), namely $[I {\cal J}^t I, G]=0$.
We still have a reduction to U$(d/2)\times$U$(d/2)$. And we still have
the two almost complex structures (we even had them in the
general case). But, since we have no hermiticity, $J_\pm$ will no longer be
hermitian with respect to {\it the same} metric $g$.

\subsection{Differential conditions}
\label{sec:diffcond}

In this subsection we discuss the differential conditions which arise both from 
 invariance of the action and from the supersymmetry algebra. We are unable to solve 
 the problem completely. The difficulties in finding
 the geneal solution may be partially ascribed to a lack of the appropriate mathematical tools. 

As discussed in subsection 6.1, 
even at the  algebraic level the natural object ${\cal J}$ does not fit into the Hitchin framework 
 unless extra restrictions are imposed, but setting\footnote{In general it is enough to 
 put $Z^t = E^{-1}T$. However for the sake of clarity we discuss only the solution $T=Z=0$.} 
 $Z=0$ and $T=0$ (as in the solution just discussed) leads to ${\cal J}$ being an almost generalized complex structure.
 We consider only the case when $T=Z=0$.

With the differential conditions the situation is very similar. 
 If we impose extra restrictions by hand then we may ensure that ${\cal J}$ is a generalized complex 
 structure. E.g., imposing $X_{\mu\nu\lambda}=0$ and $U_{\mu\nu}^{\,\,\,\,\,\,\rho}=0$ (again as in the solution above)
 we find that the conditions (\ref{SMDC3})-(\ref{SMDC5}) coincide with the conditions
 (\ref{diftopconac}), for the topological model. Therefore we can use the 
 results from Section \ref{sec:top} and conclude that the supersymmetry algebra (\ref{alfphi123}) and (\ref{alpsplus12})
  together with (\ref{SMDC3})-(\ref{SMDC5}) implies that ${\cal J}$ is a generalized complex structure. 
The remaining constraints that come from the invariance of the action 
 (\ref{SMDC1}), (\ref{SMDC2}) and (\ref{SMDC6}), can be rewritten as
\ber
\label{rest11}V_{[\nu\rho]}^{\,\,\,\,\,\,\mu} = L_{\sigma\rho} E^{\sigma\mu}_{\,\,\,\,\,\,,\nu} + 
 L_{\nu\sigma} E^{\sigma\mu}_{\,\,\,\,\,\,,\rho} - L_{\rho\nu,\sigma} E^{\sigma\mu} \\
\label{rest22} E^{\lambda\nu} V_{\nu\rho}^{\,\,\,\,\,\,\mu}  +  Y_\rho^{\,\,\,\lambda\mu} =
E^{\nu\mu}_{\,\,\,\,\,\,,\rho} J^\lambda_{\,\,\,\nu} + J^\lambda_{\,\,\,\rho,\nu} E^{\nu\mu} -
 E^{\lambda\mu}_{\,\,\,\,\,\,,\nu} J^\nu_{\,\,\,\rho} \\
 \label{rest33} Y_\nu^{\,\,\,[\lambda|\rho} E^{|\mu]\nu} =  E^{\mu\rho}_{\,\,\,\,\,\,,\nu} P^{\nu\lambda} 
 -  E^{\lambda\rho}_{\,\,\,\,\,\,,\nu} P^{\nu\mu} - E^{\nu\rho}  P^{\mu\lambda}_{\,\,\,\,\,\,,\nu}
\eer{restintact}
 and there are eight non-trivial conditions from the algebra for $S_=$, (\ref{eq:S--}). From
 (\ref{rest11})-(\ref{rest33}) we derive the following differential condition for ${\cal J}$ and
 $E$
\ber
 \nonumber E^{\lambda\nu} E^{\gamma\rho} (L_{\sigma\rho} E^{\sigma\mu}_{\,\,\,\,\,\,,\nu} + 
 L_{\nu\sigma} E^{\sigma\mu}_{\,\,\,\,\,\,,\rho} - L_{\rho\nu,\sigma} E^{\sigma\mu}) +
  E^{\gamma\mu}_{\,\,\,\,\,\,,\nu} P^{\nu\lambda} 
 -  E^{\lambda\mu}_{\,\,\,\,\,\,,\nu} P^{\nu\gamma} - E^{\nu\mu}  P^{\gamma\lambda}_{\,\,\,\,\,\,,\nu} =\\
 = E^{[\gamma|\rho} E^{\nu\mu}_{\,\,\,\,\,\,,\rho} J^{|\lambda]}_{\,\,\,\nu} + J^{[\lambda}_{\,\,\,\rho,\nu}E^{\gamma]\rho}
  E^{\nu\mu} -
 E^{[\lambda|\mu}_{\,\,\,\,\,\,,\nu} E^{|\gamma]\rho} J^\nu_{\,\,\,\rho} 
\eer{covarconstacond}
 which resembles a  condition for the complex structure to be covariantly constant.
  This is indeed the interpretation for the solutions presented below.

To summarize, the generalized sigma model (\ref{newact123}) admits (2,0) supersymmetry (\ref{neesphi}), (\ref{newsSplus})
 and (\ref{newsSminusminus}) (with $T=Z=X=U=0$) if on ${\cal M}$ there exists a generalized complex structure ${\cal J}$ 
 such (\ref{eq:stabilizer}) and a number of differential conditions is satisfied. 

Although we cannot offer an interpretation of these differential conditions in geometrical terms, 
 it is not hard  to construct additional specific examples. The main problem comes from the $S_=$ algebra. However if 
 we assume that $R$ is a complex structure, then there exists the coordinates when $R$ is constant and 
 $Y=V=0$. These assumptions do solve the $S_=$ algebra (\ref{eq:S--}), but this is not the most general solution.  
 Using this observation we may construct various examples. We start from the simplest case with a diagonal
 generalized complex structure ${\cal J}$ (i.e., $P=L=0$). In this case $J^t +R=0$ and we may use
 complex coordinates (the same for $J$ and $R$) and assume $Y=V=0$ in this coordinates. Thus the supersymmetry 
 algebra is automatically satisfied. From (\ref{eq:stabilizer}) we obtain that $E= J^t E J$ and thus $E$ is a (1,1) tensor with
 respect to $J$.
 The remaining condition (\ref{covarconstacond}) implies that $E_{i\bar{k},j} - E_{j\bar{k},i}=0$, which says that $J$ is 
 covariantly constant with respect to a connection with the torsion $H=db$.

 There exists a different way of looking for solutions. 
 We present a solution based on two reasonable assumptions.
The $N=(2,0)$ action (\ref{newact123}) has a discrete symmetry analogous to that discussed in \cite{Lindstrom:2004eh} 
 for the $N=(2,2)$ model. It is invariant under
\beq
S_{+\mu}\to -S_{+\mu}+2D_+\P^\l E_{\l\mu}\qquad S_{=\mu}\to -S_{=\mu}-2\d_=\P^\l E_{\mu\l}
\eeq{discrete}
Our first assumption is that the symmetry (\ref{discrete}) commutes with the  second supersymmetry.
 This yields nine conditions on the parameter fields, four of which are
\beq
P^{\mu\nu}=0, ~N^{~~\nu\r}_\mu =0,~Y^{~~\nu\r}_\mu =0,~U_{~~\nu\r}^\mu =0~.
\eeq{comcon}
With the additional requirement that $K=-J^t$, we solve {\em all} conditions, algebraic as well as differential. 
We find that $J$ is a complex structure which is covariantly constant with respect to the $+$-connection. 
I.e. writing $J_{\mu\nu}=J^\mu_{~\r}g_{\r\nu}$ where the hermitean metric $g_{\mu\nu}\equiv \frac 1 2 E_{(\mu\nu)}$, we have
\beq
\nabla^{(+)}_\r J_{\mu\nu}=\d_\r J_{\mu\nu}-(\Gamma^{(0)}_{\r[\mu |\tau}+H_{\r[\mu |\tau})J^\tau_{\nu]}=0~,
\eeq{covJ}
where $\Gamma^{(0)}$ is the Levi-Civita connection for $g_{\mu\nu}$ and the torsion is the three-form  $H=db$.
  The rest of
 the solution is given in terms of $J$, $E$ and $b$ according to:
\ber
&&L_{\mu\nu}=\half L_{[\mu\nu]}=J^\r_{[\nu}b_{\mu ]\r}\cr
&&R_\mu^{~\nu}=E_{\mu\r}J^\r_{~\tau}E^{\tau\nu}\cr
&&2M_{\mu\nu\r}=L_{[\nu\mu\r]}\cr
&&V_{\mu\nu}^{~~\sigma}=E_{\mu\l}\left(J^\l_{~\nu , \tau}E^{\tau\sigma}+J^\l_{~\tau}E^{\tau\sigma}_{~~,\nu}
-E^{\l\sigma}_{~~,\tau}J^\tau_{~\nu}\right)\cr
&& Q_{\mu\nu}^{~~\sigma}=J^\sigma_{~[\mu , \nu ]}~.
\eer{sln1}
In addition 
\beq
T_{\mu\nu}=0,~Z_\mu^{~\nu}=0,~ X_{\mu\nu\r}=0~.
\eeq{sln2}
This solution may  be recast in different forms using (\ref{covJ}). 
In the first example, $b \in \Omega^{1,1}({\cal M})$. Above we have analyzed the situation when  $b^{2,0}$ and $b^{0,2}$
 are allowed and 
 the generalized complex structure has the form
\beq
 {\cal J} = \left ( \begin{array}{ll} 
                   J &\,\,\,\,\, 0 \\
                   L & -J^t 
              \end{array} \right )
\eeq{formofbbL}
 where $L$ is a (2,0) and (0,2) tensor with respect to the complex structure $J$, such that $L_{ij}=2b_{ij}$.
 The metric $g$ is hermitian with respect to $J$.  
 The integrability of ${\cal J}$ implies that $\d L^{2,0}=0$. 

Analogously we consider the following generalized complex structure
\beq
 {\cal J} = \left ( \begin{array}{ll} 
                   J &\,\,\,\, P \\
                   0 & -J^t 
              \end{array} \right )
\eeq{formofbbP}
 where $P$ is a (2,0) contravariant tersor with respect to the complex structure $J$ (and $J^t +R=0$). 
 $P$ is proportional to (2,0) part of $\theta$ (antisymmetric part of $E^{-1}$). Again the differential condition
 (\ref{covarconstacond})
 can be understood as an appropriate covariantly constancy condition for $J$. 

These examples are all realized on a complex manifold ${\cal M}$. We do not know 
 if a generic solution is always a complex manifold. Notice that, in the first order model we have (incompletely) 
analyzed, there are more tensors in the game than in the second order model. Due to this, 
there are many more subcases that can be  considered.


\section{Summary}

In this paper our aim was to find a world-sheet realization of the generalized complex structure recently 
 introduced by Hitchin. We have considered three different two dimensional models inspired by the 
 first order action for the standard sigma model. The main property of these models is that the fields
 take values in $T \oplus T^*$. We have found  that the extended supersymmetry for these models is closely 
 related to the generalized complex structure. This is the main result of the paper.

We have left many  unanswered questions and open problems. E.g., we were unable to find 
 a geometrical interpretation for a generic (2,0) generalized sigma model. In general the main problem 
 is that ${\cal J}$ does not respect the natural paring on $T\oplus T^*$. Presumably one needs to introduce
  a more general Courant algebroid on $T\oplus T^*$ related to a different paring.  
  We did not consider in detail certain models which appears naturally in the present context:
 the  supersymmetric Poisson sigma model (i.e., when $E$ is Poisson 
 structure), the supersymmetric Poisson-WZ model  and the generalized sigma model with WZ term.
  Many statements from this paper can be easily extended to these models.  

Finally, and maybe most unsatisfyingly, we were not able to show the possibility of having non--complex manifolds 
as supersymmetric backgrounds for our model (which would have been a powerful motivation for the present paper), 
while not being able to rule it out either. A reason for the technical complication we are facing for the ``physical'' 
first order sigma model may be related to its non--uniqueness (i.~e., we could have taken other combinations of the 
first and second order action); maybe there are choices which make the equations simpler to solve. Independently from 
this, at the present level of development of the formalism, there are intrinsic technical difficulties coming for 
example from a big number of second and third--order tensors; presumably some better formalism to tackle with them will be needed for further progress.







{\bf Acknowledgements}:
 The work of UL is supported in part by VR grant 650-1998368.
The work of RM and AT is supported in part by EU contract 
HPRN-CT-2000-00122 and by INTAS contracts 55-1-590 and 00-0334.
We are grateful to  Marco Gualtieri,
Simeon Hellerman, Nigel Hitchin, Daniel Huybrechts and Pierre Vanhove for intersting discussions.

\appendix

\section{Appendix}
\label{appendix}

Through the paper we use the following conventions 
$$ A_{[\mu\nu]}=A_{\mu\nu}-A_{\nu\mu},\,\,\,\,\,\,\,\,
A_{(\mu\nu)}=A_{\mu\nu}+A_{\nu\mu}, \,\,\,\,\,\,\,\,
L_{[\mu\nu,\rho]}= L_{\mu\nu,\rho}+ L_{\rho\mu,\nu} + L_{\nu\rho,\mu}$$
 where $L$ is antisymmetric.

  Below we give the complete expressions for the commutators of nonmanifest 
supersymmetry acting on all fields.

\ber
\nonumber &&[\delta(\epsilon_2), \delta(\epsilon_1)]\Phi^\mu = - 2i \epsilon^+_1 \epsilon_2^+ \d_{\+} \Phi^\lambda 
 (J^\mu_{\,\,\,\nu} J^\nu_{\,\,\,\lambda} + P^{\mu\nu} L_{\nu\lambda} ) +2  \epsilon^+_1 \epsilon_2^+
 D_+ S_{+\lambda} (J^\mu_{\,\,\,\nu} P^{\nu\lambda} + P^{\mu\nu}K_\nu^{\,\,\,\lambda}) + \\
 \nonumber &&+ 2  \epsilon^+_1 \epsilon_2^+ D_+ \Phi^\lambda D_+ \Phi^\rho (J^\nu_{\,\,\,\lambda ,\rho} J^\mu_{\,\,\,\nu}
 - J^\mu_{\,\,\,\lambda,\nu} J^\nu_{\,\,\,\rho} - M_{\nu\lambda\rho}P^{\mu\nu}) +
 2 \epsilon^+_1 \epsilon_2^+ S_{+\lambda} S_{+\rho} (P^{\mu\rho}_{\,\,\,\,\,\, ,\nu} P^{\nu\lambda} -
 P^{\mu\nu} N_{\nu}^{\,\,\,\lambda\rho}) + \\
 && +  2 \epsilon^+_1 \epsilon_2^+ D_+\Phi^\nu S_{+\lambda}
 (J^\mu_{\,\,\,\nu,\rho} P^{\rho\lambda} + P^{\rho\lambda}_{\,\,\,\,\,\,,\nu} J^\mu_{\,\,\,\rho}
 - Q_{\rho\nu}^{\,\,\,\,\,\,\lambda} P^{\mu\rho} - P^{\mu\lambda}_{\,\,\,\,\,\,,\rho} J^\rho_{\,\,\,\nu})
\eer{alfphi123}
\begin{eqnarray}
\label{alpsplus12}
\nonumber&& [\delta(\epsilon_2), \delta(\epsilon_1)]S_{+\mu} = - 2i \epsilon^+_1 \epsilon_2^+ \d_{\+} S_{+\lambda}
 (K_\mu^{\,\,\,\nu} K_\nu^{\,\,\,\lambda} + L_{\mu\nu} P^{\nu\lambda}) + 2i \epsilon^+_1 \epsilon_2^+
 \d_{\+} D_+ \Phi^\lambda (K_{\mu}^{\,\,\,\nu} L_{\nu\lambda} + L_{\mu\nu} J^\nu_{\,\,\,\lambda}) + \\
\nonumber && + 2i\epsilon_1^+ \epsilon_2^+ D_+\Phi^\lambda \d_{\+} \Phi^\rho (L_{\mu\nu} J^{\nu}_{\,\,\,\lambda,\rho}
 + J^\nu_{\,\,\,\lambda} L_{\mu\rho,\nu} + K_\mu^{\,\,\,\nu} L_{\nu\rho,\lambda} + 2K_\mu^{\,\,\,\nu}
 M_{\nu\rho\lambda} - 2 J^\nu_{\,\,\,\rho} M_{\mu\nu\lambda} - Q_{\mu\lambda}^{\,\,\,\,\,\,\nu}L_{\nu\rho}) +\\
\nonumber &&+ 2i  \epsilon^+_1 \epsilon_2^+ S_{+\lambda} \d_{\+}\Phi^\rho ( - P^{\nu\lambda}_{\,\,\,\,\,\,,\rho} L_{\mu\nu}
 -L_{\mu\rho,\nu} P^{\nu\lambda} + Q_{\nu\rho}^{\,\,\,\,\,\,\lambda}K_\mu^{\,\,\,\nu} +2 L_{\nu\rho} N_{\mu}^{\,\,\,\nu\lambda}
 - J^\nu_{\,\,\,\rho} Q_{\mu\nu}^{\,\,\,\,\,\,\lambda}) + \\
\nonumber &&+ 2  \epsilon^+_1 \epsilon_2^+ D_+ S_{+\lambda} D_+\Phi^\rho (-K_{\nu\,\,\,,\rho}^{\,\,\,\lambda} K_{\mu}^{\,\,\,\nu}
 - Q_{\nu\rho}^{\,\,\,\,\,\,\lambda} K_\mu^{\,\,\,\nu} - K_{\mu\,\,\,,\nu}^{\,\,\,\lambda} J^\nu_{\,\,\,\rho} + 2P^{\nu\lambda}
 M_{\mu\nu\rho} + Q_{\mu\rho}^{\,\,\,\,\,\,\nu}K_{\nu}^{\,\,\,\lambda})+ \\
\nonumber &&+ 2 \epsilon^+_1 \epsilon_2^+ D_+\Phi^\lambda D_+\Phi^\rho D_+\Phi^\gamma (M_{\nu\lambda\rho,\gamma} 
 K_\mu^{\,\,\,\nu} - 2 J^\nu_{\,\,\,\lambda,\gamma} M_{\mu\nu\rho} + M_{\mu\lambda\rho,\nu} J^\nu_{\,\,\,\gamma}
 + Q_{\mu\rho}^{\,\,\,\,\,\,\nu} M_{\nu\lambda\gamma} ) + \\
\nonumber &&+ 2  \epsilon^+_1 \epsilon_2^+ D_+ S_{+\lambda} S_{+\rho} (2N_{\nu}^{\,\,\,\lambda\rho} K_\mu^{\,\,\,\nu}
 + K_{\mu\,\,\,,\nu}^{\,\,\,\lambda} P^{\nu\rho} - 2K_\nu^{\,\,\,\lambda} N_\mu^{\,\,\,\nu\rho} + P^{\nu\lambda}
 Q_{\mu\nu}^{\,\,\,\,\,\,\rho})+\\
\nonumber &&+ 2  \epsilon^+_1 \epsilon_2^+ S_{+\lambda} S_{+\rho} D_+\Phi^\gamma 
(N_{\nu\,\,\,\,\,\,,\gamma}^{\,\,\,\lambda\rho}
 K_{\mu}^{\,\,\,\nu} + 2 Q_{\nu\gamma}^{\,\,\,\,\,\,\lambda} N_\mu^{\,\,\,\nu\rho} + 
 N_{\mu\,\,\,\,\,\,,\nu}^{\,\,\,\lambda\rho} J^\nu_{\,\,\,\gamma} + P^{\nu\lambda}_{\,\,\,\,\,\,,\gamma} 
 Q_{\mu\nu}^{\,\,\,\,\,\,\rho} - Q_{\mu\gamma}^{\,\,\,\,\,\,\nu} N_\nu^{\,\,\,\lambda\rho} - \\
\nonumber &&- Q_{\mu\gamma\,\,\,,\nu}^{\,\,\,\,\,\,\lambda} P^{\nu\rho} ) +
 2  \epsilon^+_1 \epsilon_2^+ D_+\Phi^\rho S_{+\lambda} D_+\Phi^\gamma (
 Q_{\nu\rho\,\,\,,\gamma}^{\,\,\,\,\,\,\lambda} K_\mu^{\,\,\,\nu} - 2M_{\nu\rho\gamma} N_\mu^{\,\,\,\nu\lambda} 
 -2P^{\nu\lambda}_{\,\,\,\,\,\,,\gamma} M_{\mu\nu\rho} - M_{\mu\gamma\rho,\nu}P^{\nu\lambda} -\\
 && -J^\nu_{\,\,\,\rho,\gamma} Q_{\mu\nu}^{\,\,\,\,\,\,\lambda} + Q_{\mu\rho}^{\,\,\,\,\,\,\nu} 
 Q_{\nu\gamma}^{\,\,\,\,\,\,\lambda} + Q_{\mu\rho\,\,\,,\nu}^{\,\,\,\,\,\,\lambda} J^\nu_{\,\,\,\gamma} ) + 
 2  \epsilon^+_1 \epsilon_2^+ S_{+\lambda}S_{+\gamma} S_{+\rho} (2N_\nu^{\,\,\,\lambda\gamma}
 N_\mu^{\,\,\,\nu\rho} - N_{\mu\,\,\,\,\,\,,\nu}^{\,\,\,\lambda\gamma} P^{\nu\rho} )
\end{eqnarray}

\begin{eqnarray}
  \label{eq:S--}
 \nonumber && [\delta(\epsilon_2), \delta(\epsilon_1)]S_{=\mu} = - 2\epsilon^+_1
 \epsilon_2^+ \left[ i \d_{\+} S_{=\rho}( R_\nu^{\ \,\rho} R_\mu^{\ \,\nu}) + \right.\\
\nonumber && 
+ D_+ \d_= S_{+\rho}
     (Z_\nu^{\ \rho} R_\mu^{\ \, \nu} + K_\nu^{\ \rho} Z_\mu^{\ \nu} - 
P^{\nu\rho}T_{\mu\nu} ) + 
i\d_\+\d_= \Phi^\rho
     ( T_{\nu\rho}R_\mu^{\ \, \nu} - L_{\nu\rho} Z_\mu^{\ \nu} +
J^\nu_{\  \rho}T_{\mu\nu}) + \\
\nonumber &&
+ D_+S_{+ \nu} \d_= \Phi^\rho 
    ( U_{\sigma\rho}^{\ \ \nu}R_\mu^{\ \,\sigma} + K_{\sigma\ ,\rho}^{\ \nu}
 Z_\mu^{\ \sigma} - P^{\sigma\nu}_{\ \ \ ,\rho} T_{\mu\sigma}
+K_\sigma^{\ \nu} U_{\mu\rho}^{\ \ \sigma} -P^{\sigma\nu}X_{\mu\sigma\rho}) + \\
\nonumber && +S_{+\nu}D_+\d_=\Phi^\rho 
    ( -U_{\sigma\rho}^{\ \ \nu}R_\mu^{\ \,\sigma} + Q_{\sigma\rho}^{\ \ \nu}
 Z_\mu^{\ \,\sigma} +T_{\mu\rho,\sigma} P^{\sigma\nu} +
      P^{\sigma\nu}_{\ \ \ ,\rho} T_{\mu\sigma} +
J^\sigma_{\ \rho} U_{\mu\sigma}^{\ \ \nu} + 
T_{\sigma\rho}Y_\mu^{\ \nu\sigma}) +\\
\nonumber && + S_{+\nu} \d_= \Phi^\rho D_+ \Phi^\sigma (- U_{\lambda\rho\ \,
 ,\sigma}^{\ \ \ \nu}R_{\mu}^{\ \,\lambda}+ 
Q_{\lambda\sigma \ \, ,\rho}^{\ \ \ \nu}
 Z_\mu^{\ \,\lambda}+ X_{\mu\sigma\rho,\lambda}P^{\lambda\nu}+
 P^{\lambda\nu}_{\ \ , \rho\sigma} T_{\mu\lambda} 
+ Q_{\lambda\sigma}^{\ \ \ \nu} 
U_{\mu\rho}^{\ \ \lambda} + \\
\nonumber &&
+ U_{\mu\rho\ \, ,\lambda}^{\ \ \nu}J^\lambda_{\
 \sigma}+ J^\lambda_{\ \, \sigma, \rho} U_{\mu\lambda}^{\ \ \nu} -
 P^{\lambda\nu}_{\ \  ,\sigma}X_{\mu\lambda\rho} + P^{\lambda\nu}_{\ \ ,\rho}
 X_{\mu\sigma\lambda} + X_{\lambda\sigma\rho} Y_\mu^{\ \, \nu\lambda}
-U_{\lambda\rho}^{\ \  \nu} V_{\mu\sigma}^{\ \, \lambda}) +\\
\nonumber &&
+ i\d_\+\Phi^\nu S_{=\rho}(V_{\lambda\nu}^{\ \ \,\rho} R_\mu^{\ \,\lambda} -
 L_{\lambda\nu} Y_\mu^{\ \lambda\rho} 
+J^\sigma_{\ \, \nu} V_{\mu\sigma}^{\ \ \rho})
+ D_+ \Phi^\nu D_+S_{=\rho} ( -V_{\lambda\nu}^{\ \ \rho}  R_\mu^{\ \,\lambda} -
 R_{\mu\ ,\lambda}^{\ \,\rho}J^\lambda_{\ \, \nu}+
\\
\nonumber &&
 + R_\sigma^{\ \rho} V_{\mu\nu}^{\ \ \sigma} + R_{\sigma\ \, , \nu}^{\ \rho}
 R_\mu^{\ \sigma}) + 
D_+\Phi^\nu D_+\Phi^\rho S_{=\sigma} ( V_{\lambda\rho\ \, ,\nu}^{\ \ \,
 \sigma} R_\mu^{\ \,\lambda} +
V_{\mu\nu\ \, ,\lambda}^{\ \ \sigma} J^\lambda_{\ \ \rho} - M_{\lambda\nu\rho}
 Y_\mu^{\ \lambda\sigma} + \\ 
\nonumber &&
+J^\lambda_{\ \, \rho,\nu}V_{\mu\lambda}^{\ \
 \sigma}+V_{\lambda\rho}^{\ \ \sigma} V_{\mu\nu}^{\ \ \lambda}) 
+  D_+ S_{+\nu} S_{=\rho} (Y_\sigma^{\ \nu\rho} 
R^{\ \sigma}_\mu + K_\sigma^{\ \nu} Y_\mu^{\ \sigma\rho} -
 P^{\sigma\nu}V_{\mu\sigma}^{\ \ \rho}) +
\\
\nonumber &&
 + S_{+\nu}D_+
 S_{=\rho} (-Y_\sigma^{\ \nu\rho} R^{\ \sigma}_\mu + R_{\mu\ \, ,\lambda}^{\
 \,\rho} P^{\lambda\nu} + R_\sigma^{\ \rho} Y_\mu^{\ \nu\sigma} )
+\\
\nonumber &&
+ S_{+\nu}S_{=\rho}D_+\Phi^\sigma(- Y_{\lambda\ \  ,\sigma}^{\ \nu\rho} 
R_\mu^{\ \,\lambda} + V_{\mu\sigma\ \, ,\lambda}^{\ \ \rho} P^{\lambda\nu}
+Y_{\mu \ \ ,\lambda}^{\ \nu \rho} J^\lambda_{\ \,\sigma} +
 Q_{\lambda\sigma}^{\ \ \nu} Y_\mu^{\ \lambda\rho} + V_{\lambda\sigma}^{\ \ \rho}
 Y_\mu^{\ \nu\lambda} +P^{\lambda\nu}_{\ \, ,\sigma} V_{\mu\lambda}^{\ \ \rho} -\\
\nonumber &&
-Y_\lambda^{\ \nu\rho} V_{\mu\sigma}^{\ \ \lambda}) 
+i\d_\+ \Phi^\nu \d_= \Phi^\rho ( - L_{\sigma\nu,\rho} Z_\mu^{\ \, \sigma} + 
J^\sigma_{\ \, \nu,\rho}
T_{\mu\sigma} - L_{\sigma\nu} U_{\mu\rho}^{\ \ \sigma} + 
J^\sigma_{\ \,\nu} X_{\mu\sigma\rho} +
 X_{\sigma\nu\rho}R_\mu^{\ \sigma}) + \\
\nonumber && +\d_= S_{+\nu} S_{+\rho} (- 2 N_\sigma^{\ [\nu\rho]} L_\mu^{\ \,\sigma} -Z_{\mu\ \, ,\sigma}^{\ \,\nu}
 P^{\sigma\rho} + P^{\sigma\nu} U_{\mu\sigma}^{\ \ \rho} - Z_\sigma^{\ \nu}
 Y_\mu^{\ \rho\sigma} ) + \\
\nonumber &&
+ S_{+\nu} S_{+\rho} \d_=
 \Phi^\sigma ( - N_{\lambda \ \ ,\sigma}^{\ \nu\rho} Z_\mu^{\ \lambda} + 
U_{\mu\sigma\ \ ,\lambda}^{\ \ \ \rho} P^{\lambda\nu} -
 N_\lambda^{\ \nu\rho} U_{\mu\sigma}^{\ \ \lambda} - P^{\lambda\rho}_{\ \
 ,\sigma} U_{\mu\lambda}^{\ \ \nu}  + U_{\lambda\sigma}^{\ \ \rho}
 Y_\mu^{\ \nu\lambda}) + \\
\nonumber &&
 +  D_+ \d_= \Phi^\nu D_+ \Phi^\rho (- 2 M_{\sigma[\nu\rho]} Z_\mu^{\ \,\sigma} + T_{\mu\nu,\lambda}
 J^\lambda_{\ \ \rho} + 2 J^\lambda_{\ \, [\rho, \nu]} T_{\mu\lambda}
 -J^\sigma_{\ \, \nu} X_{\mu\rho\sigma} + X_{\sigma\rho\nu}R_\mu^{\ \sigma}
-T_{\sigma\nu}V_{\mu\rho}^{\ \ \sigma} -\\
\nonumber && 
-T_{\sigma\nu,\rho}R_\mu^{\ \sigma}) + 
 D_+\Phi^\nu D_+\Phi^\rho \d_=\Phi^\sigma (- M_{\lambda\nu\rho,\sigma}
Z_\mu^{\ \,\lambda} - X_{\mu\rho\sigma,\lambda} J^\lambda_{\ \, \nu} 
- J^\lambda_{\ \, \nu,\rho\sigma} T_{\mu\lambda} -
 M_{\lambda\nu\rho}U_{\mu\sigma}^{\ \ \lambda} - \\
\nonumber && 
-J^\lambda_{\ \, \nu,\rho} X_{\mu\lambda\sigma} + 
J^\lambda_{\ \, \rho,\sigma} X_{\mu\nu\lambda} 
+X_{\lambda\rho\sigma,\nu}R_\mu^{\ \lambda} +
 X_{\lambda\rho\sigma}V_{\mu\nu}^{\ \ \lambda})  
- D_+\Phi^\nu \d_= S_{+\rho} (Q_{\lambda\nu}^{\ \ \rho} Z_\mu^{\ \lambda} +Z_{\mu\ \, ,\lambda}^{\ \,\rho}
J^\lambda_{\ \ \nu} + \\
\nonumber && 
+P^{\lambda\rho}_{\ \ ,\nu} T_{\mu\lambda} +
 P^{\lambda\rho} X_{\mu\nu\lambda} - Z_\sigma^{\ \rho} V_{\mu\nu}^{\ \
 \sigma} 
-Z_{\sigma\ \, ,\nu}^{\ \rho}R_\mu^{\ \sigma}) + 
 S_{+\nu} S_{+\rho} S_{=\sigma}
( Y_{\mu\ \ ,\lambda}^{\ \rho\sigma}P^{\lambda\nu} -\\
\label{BIGSM} && \left.
- N_\lambda^{\ \ \nu\rho}Y_\mu^{\ \lambda\sigma} +
 Y_\lambda^{\ \rho\sigma} Y_\mu^{\ \nu\lambda}) 
\right]\ .
\end{eqnarray}


\begin{thebibliography}{6666} 
 
\newcommand{\np}{{\em Nucl.\ Phys.\ }} 
\newcommand{\pr}{{\em Phys.\ Rev.\ }} 
\newcommand{\cmp}{{\em Commun.\ Math.\ Phys.\ }} 
\newcommand{\pl}{{\em Phys.\ Lett.\ }} 
%
\bibitem{Hitchin}
N.~Hitchin,
``Generalized Calabi-Yau manifolds,''
 Q. J. Math.  {\bf 54}  (2003), no. 3, 281--308,
arXiv:math.DG/0209099.
%
\bibitem{Gualtieri}
 M.~Gualtieri,
``Generalized complex geometry,''
Oxford University DPhil thesis, arXiv:math.DG/0401221.
%
\bibitem{DH}
D.~Huybrechts,
``Generalized Calabi-Yau structures, K3 surfaces, and B-fields,''
arXiv:math.AG/0306162.
%
\bibitem{Kapustin:2000aa}
A.~Kapustin and D.~Orlov,
``Vertex algebras, mirror symmetry, and D-branes: The case of complex  tori,''
Commun.\ Math.\ Phys.\  {\bf 233} (2003) 79
[arXiv:hep-th/0010293].
%
\bibitem{fmt}
S.~Fidanza, R.~Minasian and A.~Tomasiello,
``Mirror symmetric SU(3)-structure manifolds with NS fluxes,''
arXiv:hep-th/0311122.
\bibitem{vafa}
C.~Vafa,
``Superstrings and topological strings at large N,''
J.\ Math.\ Phys.\  {\bf 42} (2001) 2798
[arXiv:hep-th/0008142].
%
\bibitem{glmw}S.~Gurrieri, J.~Louis, A.~Micu and D.~Waldram,
``Mirror symmetry in generalized Calabi-Yau compactifications,'' Nucl.\
Phys.\ B {\bf 654}, 61 (2003) [arXiv:hep-th/0211102].
%
%
\bibitem{berkovits}
N.~Berkovits,
``Super-Poincare covariant quantization of the superstring,''
JHEP {\bf 0004}, 018 (2000)
[arXiv:hep-th/0001035].
%
\bibitem{LM}
H.B. Lawson and M.L. Michelsohn, ``Spin Geometry,'' Princeton
Univ. Press, 1989.
%
\bibitem{hmw}
S.~Hellerman, J.~McGreevy and B.~Williams,
``Geometric constructions of nongeometric string theories,''
JHEP {\bf 0401} (2004) 024
[arXiv:hep-th/0208174].
%
\bibitem{Flournoy:2004vn}
A.~Flournoy, B.~Wecht and B.~Williams,
``Constructing nongeometric vacua in string theory,''
arXiv:hep-th/0404217.
%
\bibitem{Lindstrom:2004eh}
U.~Lindstrom,
``Generalized N = (2,2) supersymmetric non-linear sigma models,''
arXiv:hep-th/0401100.
%
%
\bibitem{cf}
A.~S.~Cattaneo and G.~Felder,
``A path integral approach to the Kontsevich quantization formula,''
Commun.\ Math.\ Phys.\  {\bf 212} (2000) 591
[arXiv:math.qa/9902090].
\bibitem{Gates:nk}
S.~J.~Gates, C.~M.~Hull and M.~Rocek,
``Twisted Multiplets And New Supersymmetric Nonlinear Sigma Models,''
Nucl.\ Phys.\ B {\bf 248} (1984) 157.
\bibitem{Lyakhovich:2002kc}
S.~Lyakhovich and M.~Zabzine,
``Poisson geometry of sigma models with extended supersymmetry,''
Phys.\ Lett.\ B {\bf 548} (2002) 243
[arXiv:hep-th/0210043].
%
\bibitem{hassan} 
S.~F.~Hassan,
``O(D,D:R) Deformations Of Complex Structures And Extended World
Sheet Supersymmetry,''
Nucl.\ Phys.\ B {\bf 454} (1995) 86
[arXiv:hep-th/9408060].
%

\bibitem{Lindstrom:2002jb}
U.~Lindstrom and M.~Zabzine,
``N = 2 boundary conditions for non-linear sigma models and Landau-Ginzburg
models,''
JHEP {\bf 0302} (2003) 006
[arXiv:hep-th/0209098].
%
\bibitem{Lindstrom:2002vp}
U.~Lindstrom and M.~Zabzine,
``D-branes in N = 2 WZW models,''
Phys.\ Lett.\ B {\bf 560} (2003) 108
[arXiv:hep-th/0212042].
%
%
\bibitem{kapustin}
A.~Kapustin,
``Topological strings on noncommutative manifolds,''
arXiv:hep-th/0310057.

%
\bibitem{grange}
P.~Grange,
``Branes as stable holomorphic line bundles on the non-commutative torus,''
arXiv:hep-th/0403126.


\bibitem{Baulieu:2001fi}
L.~Baulieu, A.~S.~Losev and N.~A.~Nekrasov,
``Target space symmetries in topological theories. I,''
JHEP {\bf 0202} (2002) 021
[arXiv:hep-th/0106042].
%
\bibitem{Seiberg:1999vs}
N.~Seiberg and E.~Witten,
``String theory and noncommutative geometry,''
JHEP {\bf 9909} (1999) 032
[arXiv:hep-th/9908142].
%
\bibitem{Ikeda:1993fh}
N.~Ikeda,
``Two-dimensional gravity and nonlinear gauge theory,''
Annals Phys.\  {\bf 235} (1994) 435
[arXiv:hep-th/9312059].
%
\bibitem{Schaller:1994es}
P.~Schaller and T.~Strobl,
``Poisson structure induced (topological) field theories,''
Mod.\ Phys.\ Lett.\ A {\bf 9} (1994) 3129
[arXiv:hep-th/9405110].
%
\bibitem{Klimcik:2001vg}
C.~Klimcik and T.~Strobl,
``WZW-Poisson manifolds,''
J.\ Geom.\ Phys.\  {\bf 43} (2002) 341
[arXiv:math.sg/0104189].
%


\bibitem{courant}
T.~Courant,
 ``Dirac manifolds,''
{\it  Trans. Amer. Math. Soc.}  {\bf 319}  (1990), no. 2, 631--661. 
%
\bibitem{weinstein}
T.~Courant and A.~Weinstein, 
 ``Beyond Poisson structures,'' 
 {\it Action hamiltoniennes de groupes. Troisi{\`e}me th{\'e}or{\`e}me de Lie 
 (Lyon, 1986)},  39--49, Travaux en Cours, 27, Hermann, Paris, 1988.
%

\bibitem{gilgualtieri}
G.~Cavalcanti and M.~Gualtieri, 
``Generalized complex structures on nilmanifolds,''
arXiv:math.DG/0404451.




\end{thebibliography}
\end{document}